# Direct Evidence for a Magnetic *f*-Electron Mediated Cooper Pairing Mechanism of Heavy Fermion Superconductivity in CeCoIn$_5$

J. Van Dyke[a], F. Massee[b,c], M. P. Allan[b,c,d], J. C. Davis[b,c,e,f,1], C. Petrovic[b], and D.K. Morr[a,1]

a. Department of Physics, University of Illinois at Chicago, Chicago, IL 60607, USA
b. CMPMS Department, Brookhaven National Laboratory, Upton, NY 11973, USA
c. LASSP, Department of Physics, Cornell University, Ithaca, NY 14853, USA
d. Department of Physics, ETH Zurich, CH-8093 Zurich, Switzerland
e. School of Physics and Astronomy, University of St Andrews, Fife KY16 9SS, UK
f. Kavli Institute at Cornell, Cornell University, Ithaca, NY 14853, USA

**To identify the microscopic mechanism of heavy-fermion Cooper pairing is an unresolved challenge in quantum matter studies; it may also relate closely to finding the pairing mechanism of high temperature superconductivity. Magnetically mediated Cooper pairing has long been the conjectured basis of heavy-fermion superconductivity but no direct verification of this hypothesis was achievable. Here, we use a novel approach based on precision measurements of the heavy-fermion band structure using quasiparticle interference (QPI) imaging, to reveal quantitatively the momentum-space (*k*-space) structure of the *f*-electron magnetic interactions of CeCoIn$_5$. Then, by solving the superconducting gap equations on the two heavy-fermion bands $E_k^{\alpha,\beta}$ with these magnetic interactions as mediators of the Cooper pairing, we derive a series of quantitative predictions about the superconductive state. The agreement found between these diverse predictions and the measured characteristics of superconducting CeCoIn$_5$, then provides direct evidence that the heavy-fermion Cooper pairing is indeed mediated by the *f*-electron magnetism.**

[1] To whom correspondence may be addressed. E-mail: dkmorr@uic.edu; jcseamusdavis@gmail.com

Heavy Fermions and Cooper Pairing

Superconductivity of heavy-fermions is of abiding interest, both in its own right (1-7) and because it could exemplify the unconventional Cooper pairing mechanism of the high temperature superconductors (8-11). Heavy-fermion compounds are intermetallics containing magnetic ions in the 4$f$ or 5$f$ electronic state within each unit cell. At high temperatures, each $f$-electron is localized at a magnetic ion (Fig. 1A). At low temperatures, interactions between the $f$-electron spins (red arrows Fig. 1A) lead to the formation of a narrow but subtly curved $f$-electron band $\varepsilon_k^f$ near the chemical potential (red curve, Fig. 1B), while Kondo screening hybridizes this band with the conventional $c$-electron band $\varepsilon_k^c$ of the metal (black curve, Fig. 1B). As a result, two new 'heavy-fermion' bands $E_k^{\alpha,\beta}$ (Fig. 1C) appear within a few meV of the Fermi energy. This electronic structure is controlled by the hybridization matrix element $s_k$ for inter-conversion of conduction $c$-electrons to $f$-electrons and vice-versa, such that

$$E_k^{\alpha,\beta} = \frac{\varepsilon_k^c + \varepsilon_k^f}{2} \pm \sqrt{\left(\frac{\varepsilon_k^c - \varepsilon_k^f}{2}\right)^2 + s_k^2} \qquad (1)$$

The momentum structure of the narrow bands of hybridized electronic states (Eq. 1, Fig 1C, blue curves at left) near the Fermi surface then directly reflect the form of magnetic interactions encoded within the 'parent' $f$-electron band $\varepsilon_k^f$. It is these interactions that are conjectured to drive the Cooper pairing (1-5) and thus the opening up of a superconducting energy gap (Fig. 1C, yellow curves at right).

Theoretical studies of the microscopic mechanism sustaining the Cooper pairing of such heavy-fermions typically consider the electronic fluid as a Fermi liquid but with strong antiferromagnetic spin-fluctuations derived from the $f$-electron magnetism. Moreover, it has been long hypothesized that it is these spin-fluctuations that generate the attractive Cooper pairing interaction in heavy fermion materials, specifically, in the $d$-wave channel (1-7). Why, in the decades since heavy-fermion superconductivity was discovered (12), has this been so extremely difficult to prove? The crux is that identification of the pairing mechanism in a heavy-fermion compound requires two specific pieces of information: the heavy band structures $E_k^{\alpha,\beta}$ and the $k$-space structure of the superconducting gaps $\Delta_k^{\alpha,\beta}$ (which encode the essentials of the pairing process). However, determination of the characteristic heavy-fermion band structure requires precision measurement of $E_k^{\alpha,\beta}$ both above and below the Fermi energy (Fig. 1B) making it problematic for angle resolved photoemission. Moreover, due to the extreme flatness of the heavy bands $|dE_k^{\alpha,\beta}/dk| \to 0$ and a maximum superconducting gap of typically only a few hundred $\mu$eV, no techniques existed with sufficient combined energy resolution $\delta E \leq 100 \, \mu$eV and $\boldsymbol{k}$-space resolution to directly measure $E_k^{\alpha,\beta}$ and $\Delta_k^{\alpha,\beta}$ for any heavy-fermion superconductor. Unambiguous identification of the Cooper pairing mechanism has therefore proven impossible.

Heavy-Fermion Quasiparticle Interference: Experiment and Theory

Very recently, however, this situation has changed (13-15). Heavy-fermion Bogoliubov quasiparticle interference (BQPI) imaging implemented with $\delta E \approx 75 \, \mu$eV at $T \leq 250$mK allowed detailed measurements of the $\boldsymbol{k}$-space energy gap structure $\Delta_k^{\alpha,\beta}$ for the archetypical heavy-fermion superconductor CeCoIn$_5$ (Ref.16). Its normal state properties are somewhat unconventional (17,18) and seem to reflect the presence of strong antiferromagnetic spin fluctuations (19) (whether these fluctuations are associated with the existence of a true quantum critical point (20-22) is presently unclear). The compound is electronically quasi two-dimensional (23) and its $T_c = 2.3$ K is among the highest of the heavy-fermion superconductors (16). The Cooper pairs are spin singlets (24,25) and therefore $\Delta_k^{\alpha,\beta}$ must exhibit even parity. Application of the recently developed heavy-fermion quasiparticle interference imaging technique (13,15) to CeCoIn$_5$ reveals the expected development with falling temperature of the heavy bands (26) in agreement with angle resolved photoemission (27,28). Evidence for a spin fluctuation driven pairing mechanism is adduced by comparing the change in the magnetic exchange energy to the condensation energy (29). At lower temperatures, there is clear evidence that CeCoIn$_5$ is an unconventional superconductor with a nodal energy gap (30-33) possibly with $d_{x^2-y^2}$ order parameter symmetry (15,34) (although this has not been verified directly using a phase-sensitive method (35)). The microscopic Cooper pairing mechanism of CeCoIn$_5$ has, however, not been established (7,16,36,37).

Heavy quasiparticle interference imaging and BQPI studies of CeCoIn$_5$ at $\approx$ 250 mK, can now yield accurate knowledge of the **k**-space structure of both $E_k^{\alpha,\beta}$ and $\Delta_k^{\alpha,\beta}$ (Ref. 15) Using a dilution refrigerator based SI-STM system operating down to an electron temperature of 75mK, we image the differential conductance $g(r,E)$ with atomic resolution and register, and then determine $g(q,E)$, the square-root of the power spectral density Fourier transform of each image. Recently, it was demonstrated that this approach can be used to identify elements of heavy-fermion **k**-space electronic structure (13,38) because elastic scattering of electrons from $-k(E)$ to $+k(E)$ generates density-of-states interference patterns occurring as maxima at $q(E) = k(E) - (-k(E)) = 2k(E)$ in $g(q,E)$. The onset of the heavy bands in CeCoIn$_5$ is then detected (15) as a sudden transformation of the slowly changing structure of $g(q,E)$ which appears at at E≈-4 meV, followed by a rapid evolution of the maximum intensity features towards a smaller |**q**|-radius, then by an abrupt jump to a larger |**q**|-radius, and then by a second rapid diminution of interference pattern |**q**|-radii. Thus heavy-fermion QPI reveals directly the momentum structure of two heavy bands in the energy range $-4\text{meV} < E < 12 \text{ meV}$ and the formation of a hybridization gap. Figures 2A,B are typical examples of our CeCoIn$_5$ heavy-fermion QPI data, showing the comparison of measured $g(q,E)$ and the predicted $g(q,E)$ derived from our precision model for the $E_k^{\alpha,\beta}$ (SI Section 1.2). The Fermi surface deduced from these measurements is shown in Fig. 2C, and consists of a small hole-like Fermi surface arising from the heavy $\beta$-band and two larger electron-like Fermi surfaces $(\alpha_1, \alpha_2)$ resulting from the heavy $\alpha$-band (15) (see Fig. S1). Equation 1 shows how the precise dispersions $E_k^{\alpha,\beta}$ of the two heavy-fermion bands are fixed by $\varepsilon_k^f$, which is itself generated by the Heisenberg interaction energy $I(r - r')S_r \cdot S_{r'}$, between spins $S_r$ and $S_{r'}$, at f-electron sites $r$ and $r'$ (Ref. 39) (Eq. S1, SI Section 1). Therefore, the real space (**r**-space) form of the magnetic interaction potential, $I(r - r')$, can be determined directly from the measured $E_k^{\alpha,\beta}$ (see

Eq. S2b SI Section 1). Carrying out this procedure reveals quantitatively the form of $I(\mathbf{r})$ for the $f$-electron magnetic interactions of CeCoIn$_5$ (Fig. 2D) and therefore that strong antiferromagnetic interactions occur between adjacent $f$-electron moments (SI Section 1).

Solution of Gap Equations with Magnetic $f$-electron Interaction Kernel

These interactions are hypothesized to give rise to an effective electron pairing potential $V_{SC}(\mathbf{r}-\mathbf{r}') = -I(\mathbf{r}-\mathbf{r}')/2$ (see Eq. S14, SI Section 2.1), with the opposite sign to $I(\mathbf{r}-\mathbf{r}')$ because antiparallel spins at sites $\mathbf{r}$ and $\mathbf{r}'$ (for $I > 0$) experience an attractive ($V_{SC} < 0$) pairing potential; we are assuming spin-singlet pairing throughout. Fourier transformation of $V_{SC}(\mathbf{r})$ yields $V_{SC}(\mathbf{q})$ as shown in Fig. 3A, revealing thereby that the putative pairing potential $V_{SC}(\mathbf{q})$ is strongly repulsive at $\mathbf{q} = (\pm 1, \pm 1)\pi/a_0$ and attractive at $\mathbf{q} = (\pm 1, 0)\pi/a_0; (0, \pm 1)\pi/a_0$. It is this strong repulsion at $\mathbf{q} = (\pm 1, \pm 1)\pi/a_0$ which is the long anticipated (1-11) requirement for unconventional Cooper pairing to be mediated by antiferromagnetic interactions. Finally, inserting this hypothesized pairing potential $V_{SC}(\mathbf{q})$ into the coupled superconducting gap equations for both heavy bands $E_k^{\alpha,\beta}$ of CeCoIn$_5$ yields

$$\Delta_k^{\alpha} = -\frac{x_k^2}{N}\sum_p{}' V_{SC}(\mathbf{p}-\mathbf{k})\left[x_p^2 \frac{\Delta_p^{\alpha}}{2\Omega_p^{\alpha}}\tanh\left(\frac{\Omega_p^{\alpha}}{2k_BT}\right) + w_p^2 \frac{\Delta_p^{\beta}}{2\Omega_p^{\beta}}\tanh\left(\frac{\Omega_p^{\beta}}{2k_BT}\right)\right]$$

$$\Delta_k^{\beta} = -\frac{w_k^2}{N}\sum_p{}' V_{SC}(\mathbf{p}-\mathbf{k})\left[x_{\vec{p}}^2 \frac{\Delta_{\vec{p}}^{\alpha}}{2\Omega_p^{\alpha}}\tanh\left(\frac{\Omega_p^{\alpha}}{2k_BT}\right) + w_{\vec{p}}^2 \frac{\Delta_p^{\beta}}{2\Omega_p^{\beta}}\tanh\left(\frac{\Omega_p^{\beta}}{2k_BT}\right)\right] \quad (2)$$

Here $\Omega_k^{\alpha,\beta} = \sqrt{\left(E_k^{\alpha,\beta}\right)^2 + \left(\Delta_k^{\alpha,\beta}\right)^2}$ are the two pairs of Bogoliubov bands, $w_k$ and $x_k$ are the coherence factors of the heavy fermion hybridization process, and the primed sum runs only over those momentum states $\mathbf{p}$ whose energies $E_p^{\alpha,\beta}$ lie within the interaction cutoff energy, $\omega_D$, of the Fermi energy. Our solutions to Eq. 2 (SI Section 2) predict that the $\alpha$ and $\beta$ bands of CeCoIn$_5$ possess superconducting gaps $\Delta_k^{\alpha,\beta}$ of nodal $d_{x^2-y^2}$-symmetry as shown in Fig. 3B,C. This symmetry is dictated by the large repulsive pairing potential near $\mathbf{Q} = (\pm 1, \pm 1)\pi/a_0$ (see arrow in Fig. 3A) which requires that the superconducting gap changes sign between Fermi surface points connected by $\boldsymbol{Q}$, as shown in Fig. 3B. We predict that the maximum gap value occurs on the $\alpha_1$-Fermi surface, with $\Delta_k^{\alpha,\beta}$ being well approximated by

$$\Delta_{\vec{k}}^{\alpha} = \frac{\Delta_0^{\alpha}}{2}\{[\cos(k_x) - \cos(k_y)] + \alpha_1[\cos(2k_x) - \cos(2k_y)] + \alpha_2[\cos(3k_x) - \cos(3k_y)]\}$$

$$\Delta_{\vec{k}}^{\beta} = \frac{\Delta_0^{\beta}}{2}[\cos(k_x) - \cos(k_y)]^3 \qquad (3)$$

Here $\Delta_0^{\alpha} = 0.49$ meV, $\alpha_1 = -0.61$, $\alpha_2 = -0.08$, $\Delta_0^{\beta} = -1.04$ meV represent the quantitative predictions for the $\Delta_k^{\alpha,\beta}$ in Eq. 3 derived from the hypothesis of Eq. 2 with $\omega_D = 0.66$ meV constraining the maximum possible energy gap to be $\Delta_{max} \leq 600$ μeV. The solution of Eq. 2 (without any further adjustable parameters) then also predicts $T_c = 2.96$ K, which is reduced to $T_c = 2.55$ K once one accounts for the experimentally observed mean free path of $l = 81$nm (Eqs. S21 in SI Section 2). Overall, the predicted gap structure $\Delta_k^{\alpha,\beta}$ (Eq. 3) and $T_c$ are in striking quantitative agreement with the measured $\Delta_i(\boldsymbol{k})$ (Ref. 15) and $T_c = 2.3$K (Ref. 16) for CeCoIn$_5$.

These results raise several interesting questions. First, while $\omega_D$ appears above as a phenomenological parameter, the question naturally arises of how such a crossover scale arises from the interplay between the form of the spin-excitation spectrum, the strength of the coupling between *f*-electrons and spin-fluctuations, and the flatness of the *f*-electron bands. To address this question, it will be necessary to extend the above method to a strong coupling, Eliashberg-type approach. Second, while our approach assumes that the formation of coherent (screened) Kondo lattice is concluded prior to the onset of superconductivity, it has been suggested (40) that singlet formation might continue into the superconducting state. New theoretical/experimental approaches will be required to determine if this type of composite pairing might be detectable in the temperature dependence of physical properties for $T \lesssim T_c$. Notwithstanding these questions, our first focus is now to evaluate the predictive utility of the relatively simple approach that was proposed originally[1-4] and has been implemented here.

Phase sensitive QPI

Given this detailed new understanding of $\Delta_k^{\alpha,\beta}$ (Eq. 3 and Fig. 3B,C), a variety of other testable predictions for superconducting characteristics of CeCoIn$_5$ become possible. In particular, the phase of the predicted $d_{x^2-y^2}$-symmetry gaps is directly reflected in the magnitude of $g(\boldsymbol{q}, E)$. The scattering of Bogoliubov quasiparticles between momentum points $\boldsymbol{k}_1$ and $\boldsymbol{k}_2$ near the Fermi surface leads to a contribution to $g(\boldsymbol{q} = \boldsymbol{k}_1 - \boldsymbol{k}_2, E)$ that is directly proportional to the product $\Delta_{k_1}\Delta_{k_2}$. For time-reversal invariant scalar-potential scattering, this contribution enters $g(\boldsymbol{q} = \boldsymbol{k}_1 - \boldsymbol{k}_2, E)$ with a sign that is opposite to that for time-reversal violating magnetic scattering. As a result, changes in $g(\boldsymbol{q}, E)$ generated by altering the nature of the scattering potential provide direct information on the relative phase difference between $\Delta_{k_1}$ and $\Delta_{k_2}$. This phenomenon has been beautifully demonstrated by considering magnetic field-induced changes in the conductance ratio in Bi$_2$Sr$_2$CaCu$_2$O$_8$, a single band cuprate superconductor with $d_{x^2-y^2}$-symmetry (35). Although the predicted symmetry of $\Delta_k^{\alpha,\beta}$ in CeCoIn$_5$ is similarly $d_{x^2-y^2}$ (Fig. 3), the detailed predictions for magnetic-field induced changes in $g(\boldsymbol{q}, E)$ are quite complex because of the multiple superconducting gaps and intricate band geometry (SI Section 2.3). In Fig. 4A we show our predicted values of the field induced QPI changes $\Delta g(\boldsymbol{q}, E, B) = g(\boldsymbol{q}, E, B) - g(\boldsymbol{q}, E, 0)$ for a typical energy, $E = -0.5$ meV, below the gap maximum (SI Section 2.3). For comparison, in Fig. 4B we show the measured $\Delta g(\boldsymbol{q}, E, B)$ at the same energy (SI Section 2.3). Not only is the agreement between them as to which regions of $\boldsymbol{q}$-space have enhanced or diminished scattering intensity evident, but they also demonstrate that $\Delta g(\boldsymbol{q}, E, B)$ is positive (negative) for scattering vectors, $\boldsymbol{q}_1(\boldsymbol{q}_2)$ connecting parts of the Fermi surface with the same sign (different signs) of the superconducting gap (see Fig.4C). In contrast, the equivalent predictions for phase sensitive $\Delta g(\boldsymbol{q}, E, B)$ if the gap symmetry is nodal $s$-wave, bear little discernible relationship to the experimental data (SI Section 2.3). Thus, the application in CeCoIn$_5$ of the phase sensitive QPI technique (35) reveals the predicted effects of sign changes in $d_{x^2-y^2}$ symmetry gaps of structure $\Delta_k^{\alpha,\beta}$.

Spin Excitations

Lastly, by combining the *f*-electron magnetic interactions $I(\mathbf{r})$ (Fig. 2D) and the predicted $\Delta_k^{\alpha,\beta}$ (Fig. 3), we can investigate the spin dynamics of CeCoIn$_5$ in the superconducting state (SI Section 3). One test for the nodal character of the predicted $d_{x^2-y^2}$-symmetry superconducting gap is the temperature dependence of the spin-lattice nuclear relaxation rate $1/T_1$: its theoretically predicted form is shown in Fig. 4D. This is in good agreement with that of the measured $1/T_1$ reproduced in Fig. 4E from Ref. 25. The theoretically predicted and experimentally observed power-law dependence $1/T_1 \sim T^\alpha$ with $\alpha \approx 2.5$ (straight lines in Fig. 4D and 4E) shows an exponent that is reduced from the expected $\alpha = 3$ for a single-band $d_{x^2-y^2}$-wave superconductor. We demonstrate that this effect is actually due to fine details of the $\Delta_k^{\alpha,\beta}$ multi-gap structure on the $\alpha$- and $\beta$-Fermi surfaces (see Fig. 3C). Finally, our theoretically predicted dynamic spin susceptibility in the superconducting state, using the extracted $I(\mathbf{r})$ and computed $\Delta_k^{\alpha,\beta}$ with no further adjustable parameters, exhibits a 'spin resonance' peak at $\mathbf{q} = (\pm 1, \pm 1)\pi/a_0$ and at $E \approx 0.6$ meV (SI Section 3) as shown in Fig. 4F. Such 'spin resonances' are a direct signature of an unconventional superconducting order parameter: they arise from spin-flip transitions that involve states $\mathbf{k}_1$ and $\mathbf{k}_2$ with opposite signs of the superconducting gap. The experimental data on spin excitations in superconducting CeCoIn$_5$ from inelastic neutron scattering (reproduced from Ref. 37) are shown in Fig. 4G, and exhibit a strong resonance located at $E \approx 0.6$ meV. The quantitative agreement between the predicted and measured energy of the spin resonance, as well as the form of the spin excitation spectrum both below and above $T_c$ is remarkable. Moreover, the value of $\omega_D$ used in Eq. 2 is now seen to be quite consistent with the energy scale of the spin fluctuation spectrum (SI Section S4).

Conclusions

To summarize: the $\mathbf{r}$-space and $\mathbf{q}$-space structure of magnetic interactions between *f*-electrons, $I(\mathbf{r})$ and $I(\mathbf{q})$, in the heavy-fermion state of CeCoIn$_5$ are determined quantitatively (Fig. 2D) from our measured heavy-band dispersions $E_k^{\alpha,\beta}$ (SI Section 1). The coupled superconducting gap equations (Eq. 2) are then solved using the hypothesis that it is these magnetic interactions that mediate Cooper pairing with $V_{SC}(\mathbf{q}) = -I(\mathbf{q})/2$. This allows a series of quantitative predictions regarding the physical properties of the superconducting state of CeCoIn₅. These include the superconductive critical temperature $T_c$, the **k**-space structure of the two energy gaps $\Delta_k^{\alpha,\beta}$ (Fig. 3), the phase-sensitive Bogoliubov QPI signature arising from the predicted $d_{x^2-y^2}$ symmetry of the superconducting gaps (Fig. 4A), the temperature dependence of the spin-lattice relaxation rate $1/T_1$ (Fig. 4D), and the existence and structure of a magnetic spin-resonance near $\boldsymbol{q} = (\pm 1, \pm 1)\pi/a_0$ (Fig. 4F). The demonstrated quantitative agreement between all these predictions (themselves based upon measured input parameters) and the disparate experimental characteristics of superconducting CeCoIn₅ (15,16,25,29-31,34,36,37) provide direct evidence that its Cooper pairing is indeed mediated by the residual *f*-electron magnetism (Eq. 2,3).

**Acknowledgements**

We acknowledge and thank M. Aprili, P. Coleman, M. H. Fischer, R. Flint, P. Fulde, M. Hamidian, E.-A. Kim, D.H. Lee, A.P. Mackenzie, M.R. Norman, J.P. Reid and J. Thompson for helpful discussions and communications. Supported by US DOE under contract number DEAC02-98CH10886 (J.C.D. and C.P.) and under Award No. DE-FG02-05ER46225 (D.K.M., J.V.).

**Author Contributions:** J. V. and D.K.M. performed the theoretical calculations, M.P.A. and F.M. performed the measurements and analyzed the data; C.P. synthesized the samples; D.K.M and J.C.D. supervised the investigation and wrote the paper with contributions from J.V., F.M., M.P.A., and C.P. The manuscript reflects the contributions of all authors.

**FIGURE LEGENDS**

**Figure 1 Effects of *f*-electron Magnetism in a Heavy-Fermion Material**

A. The magnetic sub-system of CeCoIn$_5$ consists of almost localized magnetic *f*-electrons (red arrows) with a weak hopping matrix element yielding a very narrow band with strong interactions between the *f*-electron spins.

B. The heavy *f*-electron band (1A) is shown schematically in red and the light *c*-electron band in black.

C. Schematic of the result of hybridizing the *c*- and *f*-electrons in B into new composite electronic states referred to as 'heavy fermions' (blue). In the right part of the panel, the opening of a superconducting energy gap is shown schematically by the back bending of the bands near the chemical potential. The microscopic interactions driving Cooper pairing of these states and thus of heavy-fermion superconductivity have not been identified unambiguously for any heavy fermion compound.

**Figure 2 Heavy-Fermion Band-structure Determination for CeCoIn$_5$**

A. Typical example of measured $g(\boldsymbol{q}, E)$ within the heavy bands.

B. Typical example of predicted $g(\boldsymbol{q}, E)$ for our parameterization of the heavy band structure. They are in good detailed agreement as are the equivalent pairs of measured and predicted $g(\boldsymbol{q}, E)$ (SI Section 1).

C. The Fermi surface of our heavy band structure model (SI Section 1).

D. $\boldsymbol{r}$-space structure of the magnetic interaction strength, $I(\boldsymbol{r})$, as obtained from Eq. S2b. This form of $I(\boldsymbol{r})$ reflects the existence of strong antiferromagnetic correlations between adjacent localized moments.

**Figure 3 Predicted Gap Structure for CeCoIn$_5$ if *f*-electron Magnetism Mediates Cooper Pairing**

A. Magnetically mediated pairing potential, $V_{sc}(\boldsymbol{q}) = -I(\boldsymbol{q})/2$, of CeCoIn$_5$. The arrow represents the momentum $\boldsymbol{Q} = (\pm 1, \pm 1)\pi/a_0$ where the pairing potential is large and repulsive.

B.  Angular dependence of the predicted superconducting gaps $\Delta_k^{\alpha,\beta}$ in the $\alpha$ and $\beta$-bands (note that the angles $\theta_\alpha$ and $\theta_\beta$ are measured around $\boldsymbol{q} = (\pi,\pi)$ and $\boldsymbol{q} = (0,0)$, respectively). The thickness and color of the Fermi surface encode the superconducting energy gap size and sign, respectively. These results were obtained at $T = 0$ using a Debye frequency of $\omega_D = 0.66$ meV. The arrow represents the momentum $\boldsymbol{Q} = (\pm 1, \pm 1)\pi/a_0$ which connects momentum states on the Fermi surfaces where the superconducting gaps possess different signs.

C.  Predicted values of the superconducting energy gaps as a function of Fermi surface angle. The largest gap is located on the $\alpha_1$-Fermi surface, whereas a small gap exists on the central $\beta$–Fermi surface and the outer $\alpha_2$-Fermi surface.

**Figure 4 Comparison of Predicted Phenomenology of CeCoIn₅ if *f*-electron Magnetism Mediates Cooper Pairing, with Experimental Data**

A  Predicted phase sensitive quasi-particle interference (PQPI) scattering pattern for the predicted $\Delta_k^{\alpha,\beta}$ with $d_{x^2-y^2}$ symmetry: $\Delta g(\boldsymbol{q}, E, B) = g(\boldsymbol{q}, E, B) - g(\boldsymbol{q}, E, 0)$ for $E = -0.5$ meV. $\Delta g(\boldsymbol{q}, E, B)$ is negative (red) for scattering vectors connecting parts of the Fermi surface with opposite signs in the order parameter, such as $\boldsymbol{q}_1$ (see panel C), while sign-preserving scattering leads to a positive $\Delta g(\boldsymbol{q}, E, B)$ (blue) such as $\boldsymbol{q}_2$ (see panel C). See SI section 2.3 for full details.

B  Measured PQPI $\Delta g(\boldsymbol{q}, E, B) = g(\boldsymbol{q}, E, B) - g(\boldsymbol{q}, E, 0)$ for $E = -0.5$ meV. The $g(\boldsymbol{q}, E, B)$ and $g(\boldsymbol{q}, E, 0)$ are measured in the identical field of view using identical measurement parameters at $B = 0$ and $B = 3T$. $\Delta g(\boldsymbol{q}, E, B)$ exhibits the same enhancement and suppression for $\boldsymbol{q}_{1,2}$ as in A. The good correspondence, especially for the relevant scattering vectors between regions whose gaps are predicted to have opposite signs, between theoretically predicted $\Delta g(\boldsymbol{q}, E, B)$ and measurements thereof is a phase sensitive verification of a $d$-wave gap symmetry in CeCoIn₅. See SI section 2.3 for additional energies and details.

C      Equal energy contour (EEC) for $E = -0.5$ meV used in panels A,B (i.e., momentum points with $E = -\Omega_{\mathbf{k}}^{\alpha\beta,}$) in the superconducting state. Scattering processes yielding the dominant contribution to $g(\mathbf{q}_{1,2}, E)$, with $\mathbf{q}_1$ ($\mathbf{q}_2$) connecting points on the EEC with opposite phases (the same phase) of the superconducting gap are shown (the phases are indicated by +/-). For simplicity, we show $\mathbf{q}'_1 = (2\pi, 2\pi) - \mathbf{q}_1$, the Umklapp-vector to $\mathbf{q}_1$. Note that the coordinate system is rotated by 45º with respect to panels A and B.

D      Predicted power law for temperature dependence of nuclear relaxation rate $1/T_1$ for the superconducting state of CeCoIn$_5$ based upon combining the *f*-electron magnetic interactions $I(\mathbf{q})$ and the predicted $\Delta_{\vec{k}}^{\alpha,\beta}$.

E      Measured temperature dependence of $1/T_1$ for the superconducting state of CeCoIn$_5$ taken from Ref.25.

F      Predicted imaginary part of the dynamical spin susceptibility $\chi(\mathbf{Q}, E)$ at $\mathbf{Q} = (\pm 1, \pm 1)\pi/a_0$ in the superconducting state of CeCoIn$_5$ based upon combining the *f*-electron magnetic interactions $I(\mathbf{q})$ and the predicted $\Delta_{\vec{k}}^{\alpha,\beta}$. A strong resonance is predicted below $T_c$ at $E \approx 0.6$ meV.

G      Measured imaginary part of the dynamical spin susceptibility $\chi(\mathbf{Q}, E)$ at $\mathbf{Q} \approx (\pm 1, \pm 1)\pi/a_0$ in the superconducting state of CeCoIn$_5$ (from Ref.37). There is a strong quantitative correspondence to the model prediction.

# Figure 1

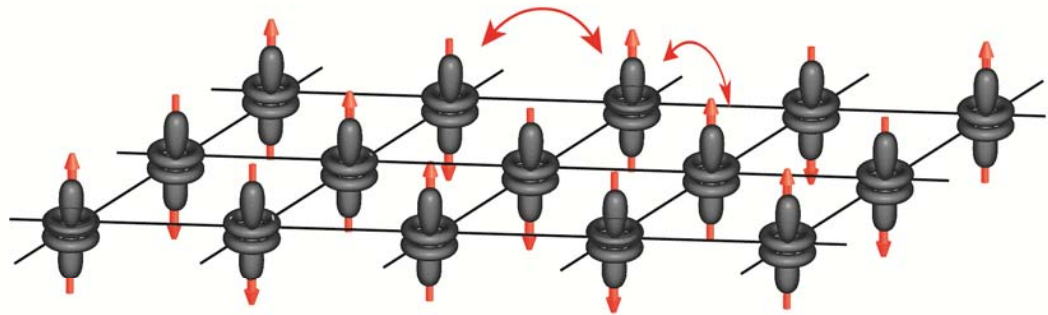

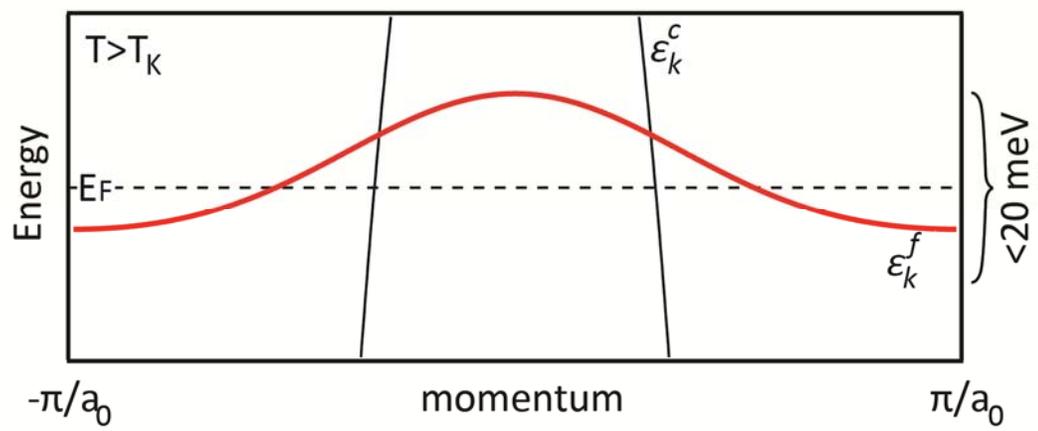

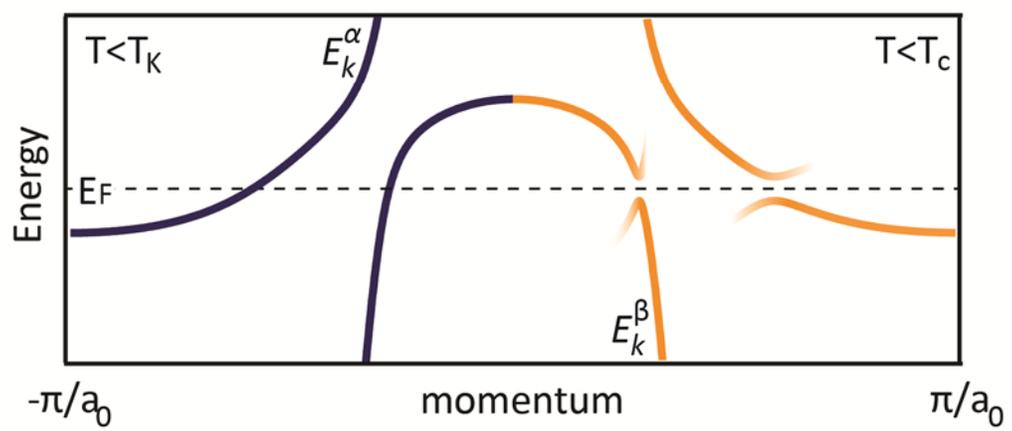

**Figure 2**

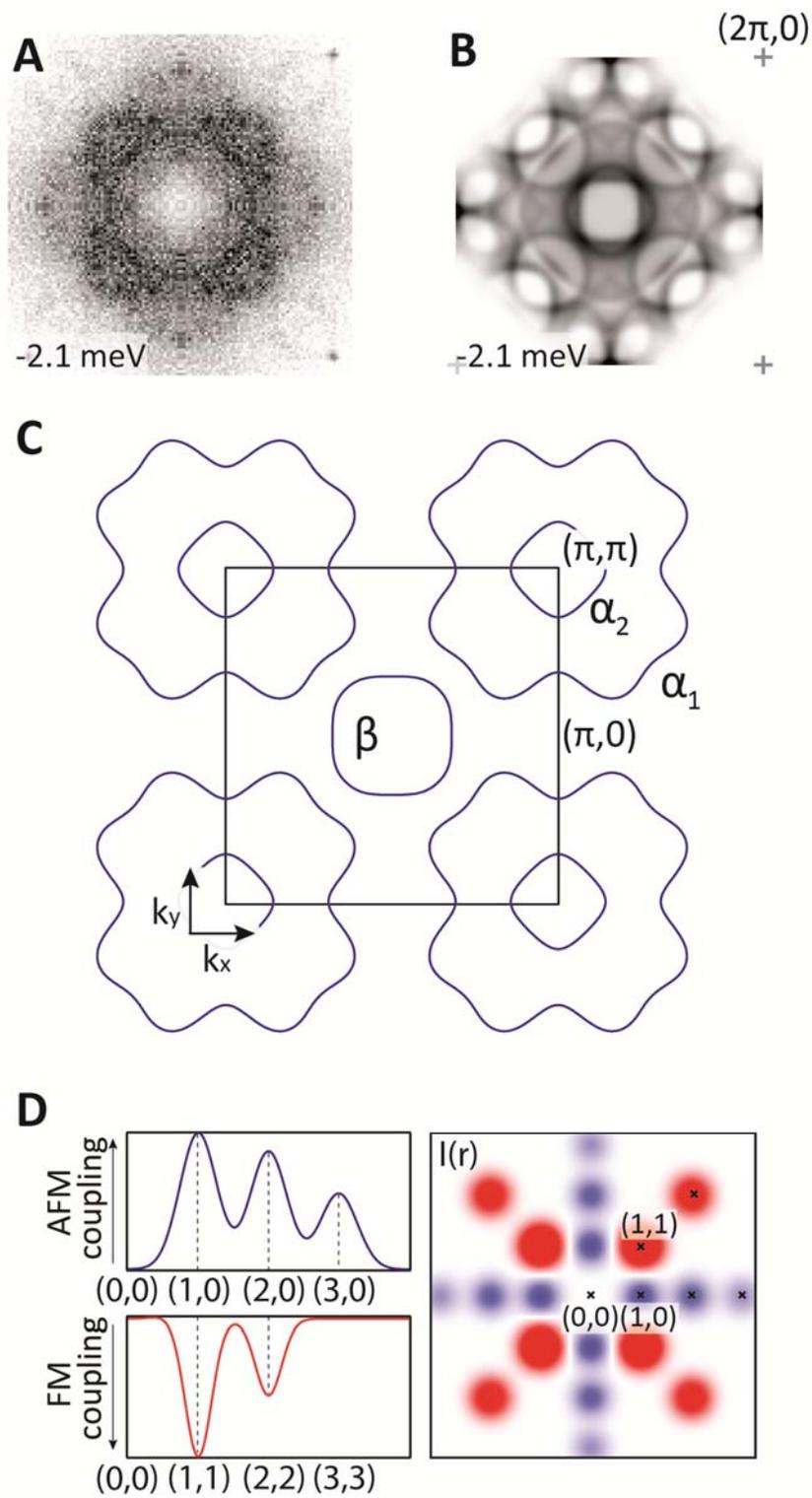

**Figure 3**

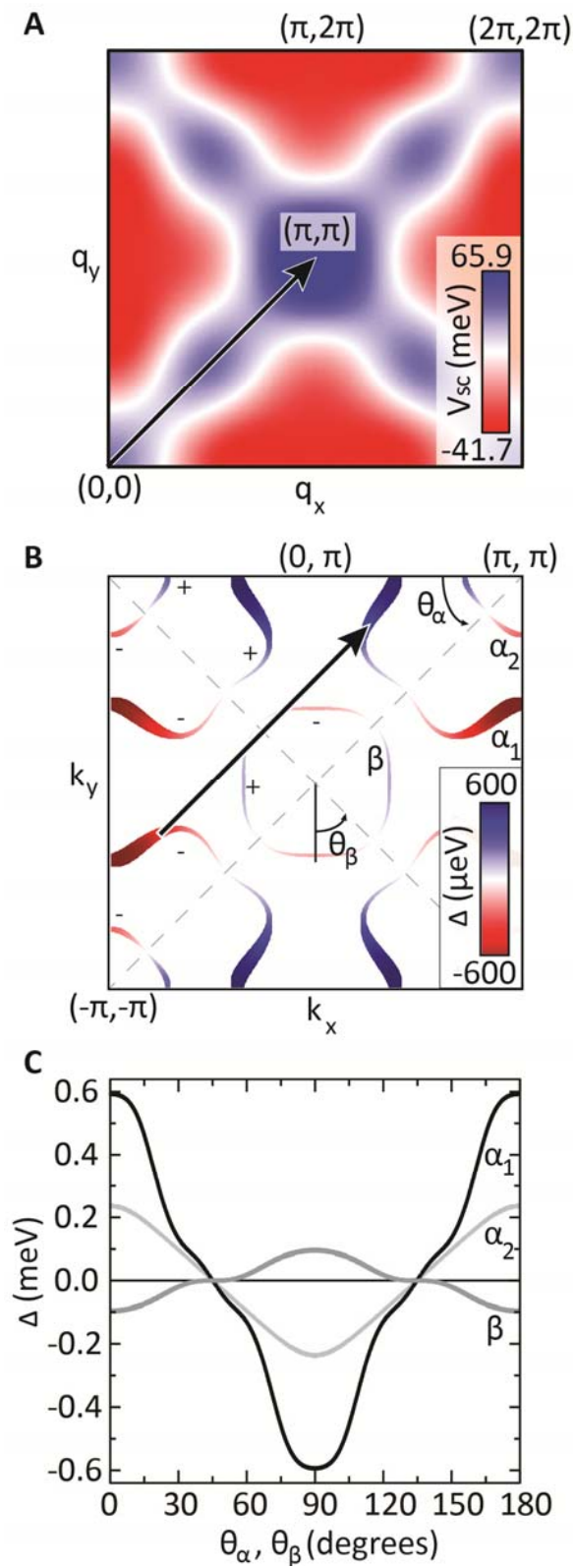

**Figure 4**

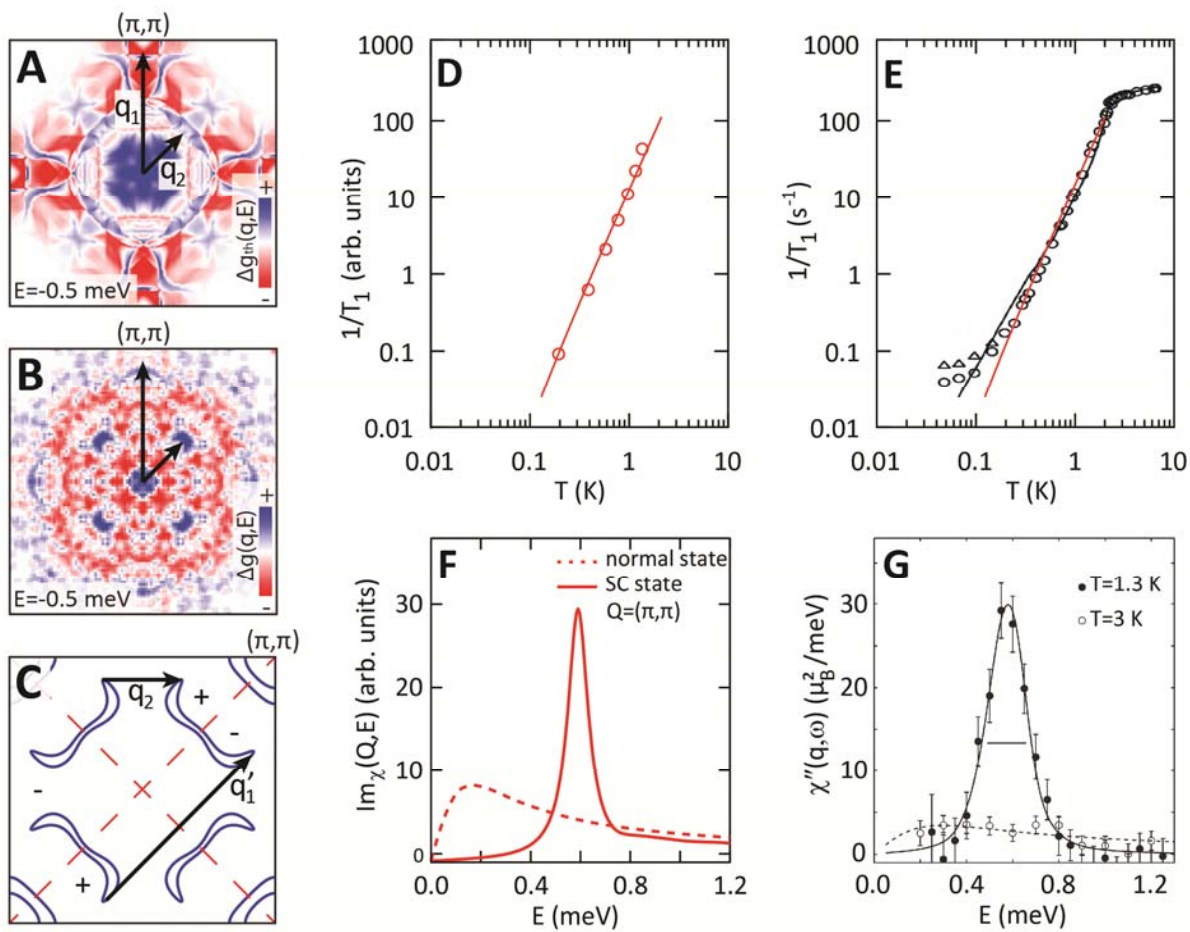

**References**

1 Miyake K, Schmitt-Rink S, Varma, CM (1986) Spin-fluctuation-mediated even-parity pairing in heavy-fermion superconductors. *Phys. Rev. B* 34(9): 6554-6556.

2 Beal-Monod MT, Bourbonnais C, Emery VJ (1986) Possible superconductivity in nearly antiferromagnetic itinerant fermion systems. *Phys. Rev. B* 34(11): 7716.

3 Scalapino DJ, Loh E, Hirsch JE (1986) *d*-wave pairing near a spin-density-wave instability. *Phys. Rev. B* 34(11): 8190-8192.

4 Scalapino DJ, Loh E, Hirsch JE (1987) Fermi-surface instabilities and superconducting *d*-wave pairing. *Phys. Rev. B* 35(13): 6694-6698.

5 Lavagna M, Millis A, Lee PA (1987) *d*-wave superconductivity in the large-degeneracy limit of the Anderson lattice. *Phys. Rev. Lett.* 58(3): 266-269.

6 Coleman P, Andrei N (1989) Kondo-stabilised spin liquids and heavy-fermion superconductivity. *J. Phys.: Condens. Matter* 1(26): 4057.

7 Monthoux P, Lonzarich G (1999) *p*-wave and *d*-wave superconductivity in quasi-two-dimensional metals. *Phys. Rev. B* 59(22): 14598-14605.

8 Monthoux P, Pines D, Lonzarich GG (2007) Superconductivity without phonons. *Nature* 450(7173): 1177-1183.

9 Norman MR (2011) The challenge of unconventional superconductivity. Science 332(6026): 196-200.

10 Scalapino DJ, (2012) A common thread: the pairing interaction for unconventional superconductors. Rev. Mod. Phys. 84(4): 1383–1417.

11 Davis JC, Lee D-H (2013) Concepts relating magnetic interactions, intertwined electronic orders and strongly correlated superconductivity. *Proc. Natl Acad Sci USA* 110(44): 17623-17630.

12 Steglich F, et al. (1979) Superconductivity in the presence of strong Pauli paramagnetism: $CeCu_2Si_2$. Phys. Rev. Lett. 43(25): 1892–1896.

13 Schmidt AR, et al. (2010) Imaging the Fano lattice to 'hidden order' transition in $URu_2Si_2$. *Nature* 465(7298): 570-576.

14 Akbari A, Thalmeier P, Eremin I (2011) RKKY interaction in the spin-density-wave phase of iron-based superconductors. *Phys Rev. B* 84(13): 134513.

15 Allan MP, et al. (2013) Imaging Cooper pairing of heavy fermions in CeCoIn5. *Nature Physics* 9(8): 468-473.

16 Petrovic C et al. (2001) Heavy-fermion superconductivity in CeCoIn5 at 2.3 K. *J. Phys.: Condens. Matter* 13(17): L337-342.

17 Kim JS, Alwood J, Stewart GR, Sarrao JL, Thompson JD (2001) Specific heat in high magnetic fields and non-Fermi-liquid behavior in CeMIn5 (M=Ir, Co). Phys. Rev. B 64(13): 134524.

18 Paglione J et al. (2003) Field-induced quantum critical point in CeCoIn$_5$. *Phys. Rev. Lett.* 91(24): 246405.

19 Hu T et al. (2012) Strong magnetic fluctuations in a superconducting state of CeCoIn$_5$. *Phys. Rev. Lett.* 108(5): 056401.

20 Howald L et al. (2011) Behavior of the quantum critical point and the Fermi-liquid domain in the heavy-fermion superconductor CeCoIn$_5$ studied by resistivity. *J. Phys. Soc. Jpn.* 80(2): 024710.

21 Paglione J, et al. (2006) Non-vanishing energy scales at the quantum critical point of CeCoIn$_5$. *Phys. Rev. Lett.* 97(10): 106606.

22 Bianchi A, Movshovich R, Vekhter I, Pagliuso PG, Sarrao JL (2003) Avoided antiferromagnetic order and quantum critical point in CeCoIn5. Phys. Rev. Lett. 91(25): 257001.

23 McCollam A, Julian SR, Rourke PMC, Aoki D, Floquet J (2005) Anomalous de Haas-van Alphen oscillations in CeCoIn5. *Phys. Rev. Lett.* 94(18): 186401.

24 Curro NJ, et al. (2012) Anomalous NMR magnetic shifts in CeCoIn$_5$. *Phys. Rev. B* 64(18): 180514.

25 Kohori Y et al. (2001) NMR and NQR studies of the heavy fermion superconductors Ce*T*In5 (*T*=Co,Ir). *Phys. Rev. B* 64(13): 134526.

26 Aynajian P, et al. (2012) Visualizing heavy fermions emerging in a quantum critical Kondo lattice. *Nature* 486(7402): 201-206.

27 Koitzsch A, et al. (2009) Electronic structure of CeCoIn5 from angle-resolved photoemission spectroscopy. *Phys. Rev. B* 79(7): 075104.

28 Jia X–W, et al. (2011) Growth characterization and Fermi surface of heavy-fermion CeCoIn$_5$ superconductor. *Chin. Phys. Lett.* 28(5): 057401.

29 Stockert O, et al. (2011) Magnetically driven superconductivity in CeCu2Si2. Nat. Phys. 7(2): 119-124.

30 Izawa K, et al. (2001) Angular position of nodes in the superconducting gap of quasi-2D heavy-fermion superconductor CeCoIn$_5$. *Phys. Rev. Lett.* 87(5): 057002.

31 Park WK, et al. (2008) Andreev reflection in the heavy-fermion superconductors and order parameter symmetry in CeCoIn$_5$. *Phys. Rev. Lett.* 100(17): 177001.

32 Ernst S, et al. (2010) Scanning tunneling microscopy studies on CeCoIn$_5$ and CeIrIn$_5$. *Phys. Status Solidi (b)* 247(3): 624–627.

33 Truncik CJS, et al. (2013) Nodal quasiparticle dynamics in the heavy fermion superconductor CeCoIn5 revealed by precision microwave spectroscopy. Nature Communications 4: 2477.

34 Zhou BB, et al. (2013) Visualizing nodal heavy fermion superconductivity in CeCoIn$_5$. *Nature Physics* **9**, 474.

35 Hanaguri T, et al. (2009) Coherence factors in a high-$T_c$ cuprate probed by quasi-particle scattering off vortices. *Science* 323(5926): 923-926.

36 Kohori Y et al. (2001) NMR and NQR studies of the heavy fermion superconductors Ce*T*In5 (*T*=Co,Ir). *Phys. Rev. B* 64(13): 134526.

37 Stock C, et al. (2008) Spin resonance in the *d*-wave superconductor CeCoIn5. *Phys. Rev. Lett.* 100(8): 087001.

38 Yuan T, Figgins J, and Morr DK (2012) Hidden Order Transition in URu$_2$Si$_2$: Evidence for the Emergence of a Coherent Anderson Lattice from Scanning Tunneling Spectroscopy. *Phys. Rev. B* 86(3): 035129

39 Senthil T, Vojta M, Sachdev S (2004) Weak magnetism and non-Fermi liquids near heavy-fermion critical points. *Phys. Rev. B* 69(3): 035111.

40 Flint R, Nevidomskyy AH, Coleman P (2011) Composite Pairing in a mixed-valent two-channel Anderson model. Phys. Rev. B 84(6): 064514.

# Supplementary Information:

# Direct Evidence for a Magnetic *f*-Electron Mediated Cooper Pairing Mechanism of Heavy Fermion Superconductivity in CeCoIn$_5$

J. Van Dyke, F. Massee, M. P. Allan, C. Petrovic, J. C. Davis, and D.K. Morr

## 0. Outline

Before describing the theoretical elements of our approach in the following sections, we briefly summarize it here. Our microscopic model for the description of heavy fermion materials in the normal (hybridized) state is described in Sec.1.1. The hybridized band structure of CeCoIn$_5$ in the normal state is extracted from scanning tunneling microscopy (STM) data via heavy quasi-particle interference scattering, as described in Sec.1.2 [see Eq. S8 and Fig. S1]. Since, within the model, the dispersion of the heavy band is directly linked to the strength of the *f*-electron magnetic interactions in real space, $I_{\mathbf{r},\mathbf{r}'}$ [see Eq. S2b], we can determine the latter from the measured hybridized band structure [see Eq. S13]. Starting from the proposal that these magnetic interactions act as the superconducting pairing interaction, we solve the multi-band BCS gap equations [see Eq. S20] and predict the multiband superconducting gap structure, all with a $d_{x^2-y^2}$-symmetry (see Sec.2.1). Using the predicted SC gap structure, we then compute the Bogoliubov quasi particle interference (BQPI) pattern [see Sec.2.2, Eq. S25] , as well as the phase-sensitive quasi particle interference (PQPI) pattern [see Sec.2.3, Eq. S32]. Finally, using the magnetic interactions as well as the predicted superconducting gap, we predict the emergence of a resonance peak in superconducting state [see Sec. S3, Eq. S45], and the temperature dependence of the spin lattice relaxation rate, $1/T_1$ [see Eq. S47]. Finally we show that all of these predictions from our theory are in good quantitative agreement with the experimental data, as discussed in more detail below.

# 1. Heavy Fermions in the Normal State

## 1.1 The Hamiltonian, Mean-Field Equations and Energy Dispersions

To understand the heavy-fermion electronic structure of CeCoIn5, we use a model based upon the periodic Anderson lattice in the infinite $U$ limit. To this end, we employ the slave boson approach (1-4) for which the Hamiltonian is given by

$$H = \sum_{\mathbf{k},\sigma} \varepsilon_{\mathbf{k}}^c c_{\mathbf{k}\sigma}^\dagger c_{\mathbf{k}\sigma} + E_0 \sum_{\mathbf{r}} n_{\mathbf{r}}^f + \sum_{\mathbf{r},\mathbf{r}',\sigma} V_{\mathbf{r},\mathbf{r}'} f_{\mathbf{r},\sigma}^\dagger b_{\mathbf{r}} c_{\mathbf{r}',\sigma} + h.c. + \sum_{\mathbf{r},\mathbf{r}'} I_{\mathbf{r},\mathbf{r}'} \mathbf{S}_{\mathbf{r}} \cdot \mathbf{S}_{\mathbf{r}'}. \tag{S1}$$

Here $c_{\mathbf{k},\sigma}^\dagger (f_{\mathbf{k},\sigma}^\dagger)$ creates an electron with spin $\sigma$ and momentum $\mathbf{k}$ in the light conduction $c$-band (magnetic $f$-band) whose two-dimensional (2D) dispersion is $\varepsilon_{\mathbf{k}}^c$ ($\varepsilon_{\mathbf{k}}^f$). $V_{\mathbf{r},\mathbf{r}'}$ is the hybridization matrix element between site $\mathbf{r}$ in the $f$-band and site $\mathbf{r}'$ in the $c$-band. The $b_{\mathbf{r}}^\dagger, b_{\mathbf{r}}$ are slave-boson operators, that account for fluctuations between unoccupied and singly occupied $f$-electron sites via the constraint $\sum_\alpha f_{\mathbf{r},\alpha}^\dagger f_{\mathbf{r},\alpha} + b_{\mathbf{r}}^\dagger b_{\mathbf{r}} = 1$ (1-3). $I_{\mathbf{r},\mathbf{r}'}$ is the magnetic interaction strength in the $f$-band between moment locations $\mathbf{r}, \mathbf{r}'$, and described by the $S = 1/2$ spin operator $\mathbf{S}_{\mathbf{r}}$ (here, the sum runs over all different pairs $\mathbf{r}, \mathbf{r}'$). In a path integral approach, one employs a fermionic SU(2) representation of the spin operator via $\mathbf{S}_{\mathbf{r}} = \frac{1}{2} \sum_{\alpha,\beta} f_\alpha^\dagger \boldsymbol{\sigma}_{\alpha\beta} f_\beta$ where $\boldsymbol{\sigma}_{\alpha\beta} = \left( \sigma_{\alpha\beta}^x, \sigma_{\alpha\beta}^y, \sigma_{\alpha\beta}^z \right)$ is a vector of Pauli-matrices (3,5-7). The magnetic interaction term is then decoupled using a Hubbard-Stratonovich field, $t_f(\mathbf{r}, \mathbf{r}', \tau)$, and the constraint is enforced by means of a Lagrange multiplier $(\varepsilon_f - E_0)$. In the static approximation, one replaces, $b_{\mathbf{r}}^\dagger, b_{\mathbf{r}}$ by their (real) expectation value $z_0$, and $t_f(\mathbf{r}, \mathbf{r}', \tau)$ by its static expectation value $t_f(\mathbf{r}, \mathbf{r}')$. Minimizing the effective action then leads to the self-consistent equations (8)

$$s(\mathbf{r}, \mathbf{r}') = \frac{J_{\mathbf{r},\mathbf{r}'}}{\pi} \int_{-\infty}^{\infty} d\omega \, n_F(\omega) \, \mathrm{Im} G_{fc}(\mathbf{r}, \mathbf{r}', \omega) \tag{S2a}$$

$$t_f(\mathbf{r}, \mathbf{r}') = -\frac{I_{\mathbf{r},\mathbf{r}'}}{\pi} \int_{-\infty}^{\infty} d\omega \, n_F(\omega) \, \mathrm{Im} G_{ff}(\mathbf{r}, \mathbf{r}', \omega) \tag{S2b}$$

$$n_f(\mathbf{r}) = -\int_{-\infty}^{\infty} \frac{d\omega}{\pi} \, n_F(\omega) \, \mathrm{Im} G_{ff}(\mathbf{r}, \mathbf{r}, \omega) \tag{S2c}$$

where $n_f(\mathbf{r}) = 1 - z_0^2(\mathbf{r})$ and $J_{\mathbf{r},\mathbf{r}'} = V_{\mathbf{r},\mathbf{r}'}^2 / (\varepsilon_f - E_0) > 0$. The effective hybridization $s(\mathbf{r}, \mathbf{r}') = V_{\mathbf{r},\mathbf{r}'} z_0(\mathbf{r})$ represents the screening of a magnetic moment while $t_f(\mathbf{r}, \mathbf{r}')$ describes the antiferromagnetic correlations between moments. For a translationally invariant system, $s(\mathbf{r},\mathbf{r}') = s(\mathbf{r}-\mathbf{r}')$, $t_f(\mathbf{r},\mathbf{r}') = t_f(\mathbf{r}-\mathbf{r}')$, and $I_{\mathbf{r},\mathbf{r}'} = I(\mathbf{r}-\mathbf{r}')$. Here, $t_f(\mathbf{r}-\mathbf{r}')$ are the effective hopping matrix elements for $f$-electrons hopping between ($f$-electron) sites $\mathbf{r}$ and $\mathbf{r}'$ giving rise to a dispersion, $\varepsilon_\mathbf{k}^f$, of the 'parent' $f$-electron band [see discussion below Eq. S11]. In momentum space the renormalized quasi-particle Greens functions $G_{cc}$, $G_{ff}$ and $G_{fc}$, describing the hybridization process, are given by

$$G_{cc}(\mathbf{k},\sigma,\omega) = \frac{w_\mathbf{k}^2}{\omega - E_\mathbf{k}^\alpha + i\Gamma} + \frac{x_\mathbf{k}^2}{\omega - E_\mathbf{k}^\beta + i\Gamma} \tag{S3a}$$

$$G_{ff}(\mathbf{k},\sigma,\omega) = \frac{x_\mathbf{k}^2}{\omega - E_\mathbf{k}^\alpha + i\Gamma} + \frac{w_\mathbf{k}^2}{\omega - E_\mathbf{k}^\beta + i\Gamma} \tag{S3b}$$

$$G_{cf}(\mathbf{k},\sigma,\omega) = w_\mathbf{k} x_\mathbf{k} \left[ \frac{1}{\omega - E_\mathbf{k}^\alpha + i\Gamma} - \frac{1}{\omega - E_\mathbf{k}^\beta + i\Gamma} \right] \tag{S3c}$$

with $\Gamma$ being the inverse lifetime of the quasi-particle states, and heavy fermion coherence factors $w_\mathbf{k}, x_\mathbf{k}$ given by

$$w_\mathbf{k}^2, x_\mathbf{k}^2 = \frac{1}{2}\left[1 \pm \frac{\left(\frac{\varepsilon_\mathbf{k}^c - \varepsilon_\mathbf{k}^f}{2}\right)^2}{\sqrt{\left(\frac{\varepsilon_\mathbf{k}^c - \varepsilon_\mathbf{k}^f}{2}\right)^2 + s_\mathbf{k}^2}}\right] \tag{S4a}$$

$$w_\mathbf{k} x_\mathbf{k} = \frac{s_\mathbf{k}}{2\sqrt{\left(\frac{\varepsilon_\mathbf{k}^c - \varepsilon_\mathbf{k}^f}{2}\right)^2 + s_\mathbf{k}^2}} \tag{S4b}$$

and, most importantly,

$$E_\mathbf{k}^{\alpha,\beta} = \frac{\varepsilon_\mathbf{k}^c + \varepsilon_\mathbf{k}^f}{2} \pm \sqrt{\left(\frac{\varepsilon_\mathbf{k}^c - \varepsilon_\mathbf{k}^f}{2}\right)^2 + s_\mathbf{k}^2} \tag{S4c}$$

are the renormalized band dispersions reflecting the hybridization between the $c$- and $f$-bands (this is Eq. 1 of the main text).

Equivalent to the static approximation is a mean-field decoupling on the Hamiltonian level (with the same mean-field equations as given above) leading to the effective mean-field Hamiltonian

$$H_{MF} = \sum_{\mathbf{k},\sigma} \varepsilon_{\mathbf{k}}^c c_{\mathbf{k},\sigma}^\dagger c_{\mathbf{k},\sigma} + \sum_{\mathbf{k}} \varepsilon_{\mathbf{k}}^f f_{\mathbf{k},\sigma}^\dagger f_{\mathbf{k},\sigma} + \sum_{\mathbf{k},\sigma} s_{\mathbf{k}} f_{\mathbf{k},\sigma}^\dagger c_{\mathbf{k},\sigma} + h.c. \tag{S5}$$

in which the conventional metallic $c$-band states are hybridized with the magnetic $f$-electron states via an effective hybridization matrix element $s_{\mathbf{k}}$. Diagonalizing the mean-field Hamiltonian using the unitary transformation

$$c_{\mathbf{k},\sigma}^\dagger = w_{\mathbf{k}} \alpha_{\mathbf{k},\sigma}^\dagger + x_{\mathbf{k}} \beta_{\mathbf{k},\sigma}^\dagger \tag{S6a}$$

$$f_{\mathbf{k},\sigma}^\dagger = -x_{\mathbf{k}} \alpha_{\mathbf{k},\sigma}^\dagger + w_{\mathbf{k}} \beta_{\mathbf{k},\sigma}^\dagger \tag{S6b}$$

with $w_{\mathbf{k}}, x_{\mathbf{k}}$ given above, then yields

$$H_{MF} = \sum_{\mathbf{k},\sigma} \left( E_{\mathbf{k}}^\alpha \alpha_{\mathbf{k},\sigma}^\dagger \alpha_{\mathbf{k},\sigma} + E_{\mathbf{k}}^\beta \beta_{\mathbf{k},\sigma}^\dagger \beta_{\mathbf{k},\sigma} \right) \tag{S7}$$

## 1.2 Heavy Quasiparticle Interference (HQPI) in the Normal State of CeCoIn$_5$

In the presence of impurities, the STM tunneling conductance $dI/dV$ varies spatially, and the QPI spectrum can be measured by taking the square root of the power spectral density, $g(\mathbf{q}, E)$, of $dI/dV(\mathbf{r}, E)$ into $\mathbf{q}$-space. Here, $g(\mathbf{q}, E)$ is the magnitude of the conventional Fourier transform of $dI/dV(\mathbf{r}, E)$ into $\mathbf{q}$-space, the latter of which we denote by $\bar{g}(\mathbf{q}, E)$, i.e., $g(\mathbf{q}, E) = |\bar{g}(\mathbf{q}, E)|$. In the Born approximation, we then obtain for each of the spin projections

$$\bar{g}(\mathbf{q}, E) \equiv \frac{dI(\mathbf{q}, E)}{dV} = \frac{\pi e}{\hbar} N_t \sum_{i,j=1}^{2} \left[ \hat{t} \hat{N}(\mathbf{q}, E) \hat{t} \right]_{ij} \tag{S8a}$$

$$\hat{N}(\mathbf{q}, E) = -\frac{1}{\pi} \text{Im} \int \frac{d^2k}{(2\pi)^2} \hat{G}(\mathbf{k}, E) \hat{U} \hat{G}(\mathbf{k} + \mathbf{q}, E) \tag{S8b}$$

where

$$\hat{U} = \begin{pmatrix} U_{cc} & U_{cf} \\ U_{fc} & U_{ff} \end{pmatrix} \tag{S9a}$$

$$\hat{t} = \begin{pmatrix} -t_c & 0 \\ 0 & -t_f \end{pmatrix} \tag{S9b}$$

and

$$\hat{G}(\mathbf{k}, E) = \begin{pmatrix} G_{cc}(\mathbf{k}, \sigma, E) & G_{cf}(\mathbf{k}, \sigma, E) \\ G_{fc}(\mathbf{k}, \sigma, E) & G_{ff}(\mathbf{k}, \sigma, E) \end{pmatrix} \tag{S10}$$

Here, $N_t$ is the density of states in the STM tip, and, $t_c$ and $t_f = t_f^{(0)} z_0$ are the amplitudes for electron tunneling from the STM tip into the light and heavy bands (10), respectively. Moreover, $U_{cc}$ and $U_{ff} = U_{ff}^{(0)} z_0^2$ are the scattering potential for intraband scattering in the $c$- and $f$-electron bands, respectively, while $U_{fc} = U_{cf} = U_{cf}^{(0)} z_0$ is the scattering potential for interband scattering between the $c$- and $f$-electron bands. For the theoretical results of $g(\mathbf{q}, E)$ shown in the main text and below, we used $t_f/t_c = 0.05$, $U_{ff}/U_{cc} = 0.15$ and $U_{fc}/U_{cc} = 0.16$.

The present best fit of the theoretical heavy QPI spectrum Eq. S8a using the band structure of Eq. S4c, to the experimental $g(\mathbf{q}, E)$ data as shown in Fig. S1 then yields $c$-electron and $f$-electron band structures separately

$$\varepsilon_{\mathbf{k}}^c = -2t_{c1}[\cos(k_x) + \cos(k_y)] - 4t_{c2}\cos(k_x)\cos(k_y) - 2t_{c3}[\cos(2k_x) + \cos(2k_y)] - \mu_c \tag{S11a}$$

$$\varepsilon_{\mathbf{k}}^f = -2t_{f1}[\cos(k_x) + \cos(k_y)]$$
$$- 4t_{f2}\cos(k_x)\cos(k_y) - 2t_{f3}[\cos(2k_x) + \cos(2k_y)]$$
$$-4t_{f5}\cos(2k_x)\cos(2k_y) - 2t_{f7}[\cos(3k_x) + \cos(3k_y)] + \varepsilon_f \tag{S11b}$$

with $t_{c1} = -50.0$ meV, $t_{c2} = -13.36$ meV, $t_{c3} = 16.73$ meV, $\mu_c = 151.51$ meV, $t_{f1} = -0.85$ meV, $t_{f2} = -0.35$ meV, $t_{f3} = -0.8$ meV, $t_{f5} = 0.1$ meV, $t_{f7} = 0.09$ meV, $\varepsilon_f =$

0.5 meV. Here, $t_{fi}$ is related to the hopping matrix element $t_f(\mathbf{r} - \mathbf{r}')$ introduced in Eq. S2 as follows: $t_f(\mathbf{r} - \mathbf{r}') = t_{f1}$ for $\mathbf{r} - \mathbf{r}' = (\pm 1, 0)a_0$ or $(0, \pm 1)a_0$, $t_f(\mathbf{r} - \mathbf{r}') = t_{f2}$ for $\mathbf{r} - \mathbf{r}' = (\pm 1, \pm 1)a_0$, $t_f(\mathbf{r} - \mathbf{r}') = t_{f3}$ for $\mathbf{r} - \mathbf{r}' = (\pm 2, 0)a_0$ or $(0, \pm 2)a_0$, $t_f(\mathbf{r} - \mathbf{r}') = t_{f5}$ for $\mathbf{r} - \mathbf{r}' = (\pm 2, \pm 2)a_0$, and $t_f(\mathbf{r} - \mathbf{r}') = t_{f7}$ for $\mathbf{r} - \mathbf{r}' = (\pm 3, 0)a_0$ or $(0, \pm 3)a_0$. Thus, $t_{f1}, t_{f2},$ etc. are the matrix elements for nearest-neighbor, next-nearest-neighbor, etc. hopping.

Moreover, to achieve this good fit we find that the **k**-space structure of the hybridization process is given by

$$s_\mathbf{k} = s_0 + s_1 [\sin(k_x) \sin(k_y)]^2 \qquad (S12)$$

with $s_0 = 3$ meV, and $s_1 = 7$ meV. In Figs. S1(a) and (b), we present a comparison of the experimentally measured and theoretically computed HQPI dispersions. We note that for this set of parameters, the $f$-electron occupation per site $n_f(\mathbf{r}) = 0.85$ which is in very good agreement with that obtained in recent x-ray absorption near-edge structure studies (9).

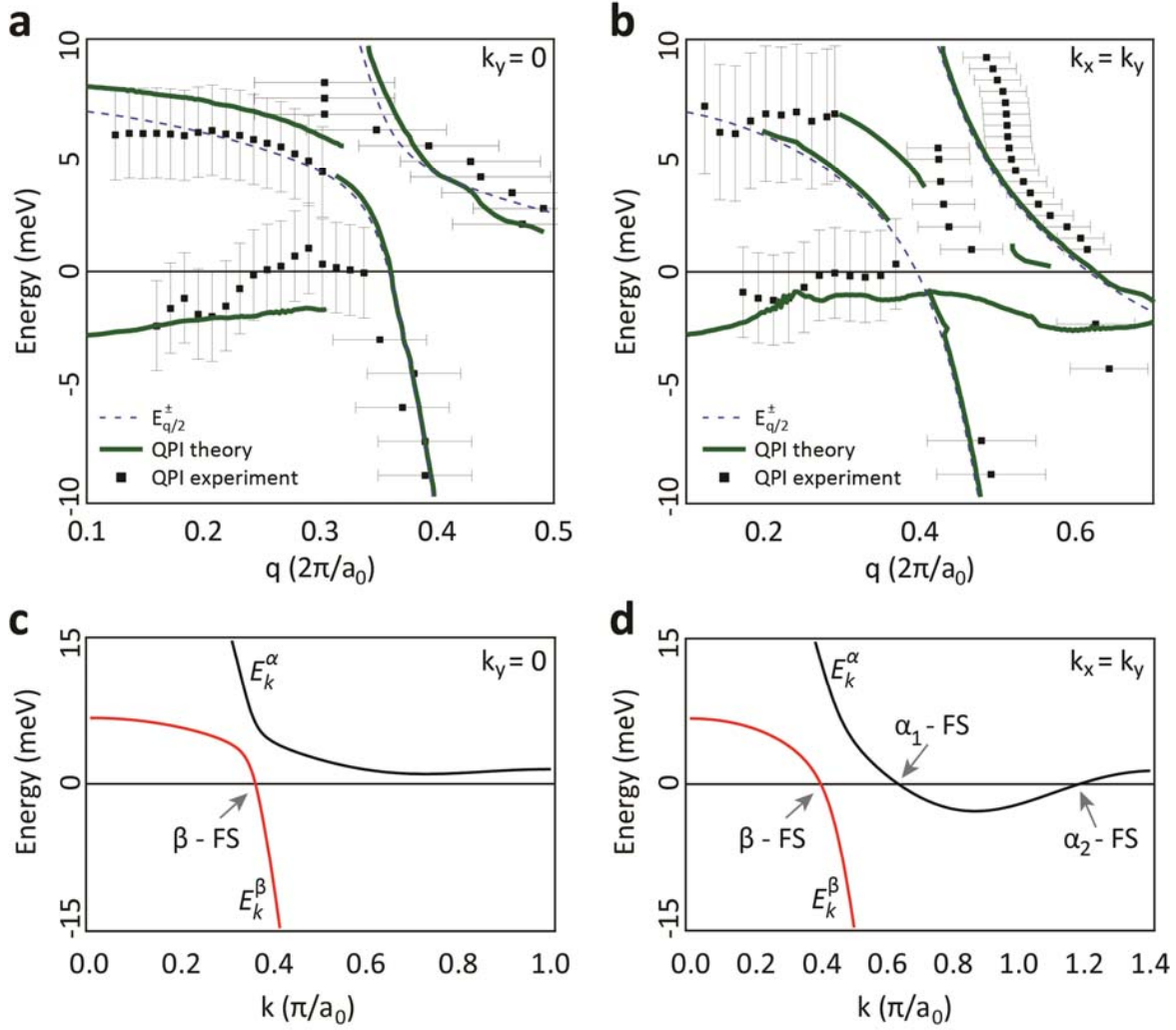

**Fig. S1.** Comparison between the dispersions of maxima in the experimental data and that in the theoretically computed $g(\mathbf{q}, E)$ for the (0,1) (**a**) and (1,1) (**b**) directions. The black dots mark the positions of maxima extracted from measured $g(\mathbf{q}, E)$ layers; the solid lines mark the positions of maxima extracted from theoretically computed $g(\mathbf{q}, E)$ cuts; the dashed lines are the expected HQPI dispersions arising from $2k_F$-scattering (see Ref. *10*). (c), (d) $E_{\mathbf{k}}^{\alpha,\beta}$ along $k_y = 0$ and $k_x = k_y$.

To obtain these fits for $\varepsilon_{\mathbf{k}}^c, \varepsilon_{\mathbf{k}}^f$, and $s_{\mathbf{k}}$, we first note that for energies sufficiently removed from the Fermi energy, one has $E_{\mathbf{k}}^{\alpha,\beta} \approx \varepsilon_{\mathbf{k}}^c$ and the QPI spectrum is thus determined predominantly by the light band. Fitting the experimental QPI spectrum for these energies first thus allows us to obtain $\varepsilon_{\mathbf{k}}^c$. For energies close to the Fermi energy, the available experimental QPI data provide a sufficient number of constraints to independently determine both $\varepsilon_{\mathbf{k}}^f$ and $s_{\mathbf{k}}$ through a comparison with the theoretically predicted QPI spectrum for the resulting $E_{\mathbf{k}}^{\alpha,\beta}$. In Figs. S1(c) and (d), we also present $E_{\mathbf{k}}^{\alpha,\beta}$ along $k_y = 0$ and $k_x = k_y$, respectively, where we have indicated the single Fermi surface crossing of the $\beta$-band, and the two Fermi surface crossings of the $\alpha$-band, denoted by $\alpha_1$ and $\alpha_2$.

Within the context of Eq. S2b the dispersion of the heavy $f$-band is directly linked to the strength of the magnetic interaction, $I_{\mathbf{r},\mathbf{r}'}$. Next we use the quantitative results in Eq. S11 and Eq. S12 to obtain the normal state Greens functions in Eq. S3 and to subsequently solve Eq. S2b for the magnetic interactions, $I_{\mathbf{r},\mathbf{r}'} = I(\mathbf{r} - \mathbf{r}')$ in real space. By using this approach, we obtain $I(\mathbf{r} - \mathbf{r}') = I_1 = 6.44$ meV for $\mathbf{r} - \mathbf{r}' = (\pm 1, 0)a_0$ or $(0, \pm 1)a_0$, $I(\mathbf{r} - \mathbf{r}') = I_2 = -20.30$ meV for $\mathbf{r} - \mathbf{r}' = (\pm 1, \pm 1)a_0$, $I(\mathbf{r} - \mathbf{r}') = I_3 = 6.04$ meV for $\mathbf{r} - \mathbf{r}' = (\pm 2, 0)a_0$ or $(0, \pm 2)a_0$, $I(\mathbf{r} - \mathbf{r}') = I_5 = -9.65$ meV for $\mathbf{r} - \mathbf{r}' = (\pm 2, \pm 2)a_0$, and $I(\mathbf{r} - \mathbf{r}') = I_7 = 2.58$ meV for $\mathbf{r} - \mathbf{r}' = (\pm 3, 0)a_0$ or $(0, \pm 3)a_0$. Here, a positive (negative) value of $I(\mathbf{r} - \mathbf{r}')$ represents antiferromagnetic (ferromagnetic) coupling between spins located at $\mathbf{r}, \mathbf{r}'$. The resulting real space structure of $I(\mathbf{r})$ is shown in Fig. 2D of the main text, leading to antiferromagnetic correlations between adjacent localized moments

The momentum space structure of $f$-electron magnetism in CeCoIn$_5$, i.e., the Fourier transform of $I(\mathbf{r})$, is then given by

$$I(\mathbf{q}) = 2I_1[\cos(q_x) + \cos(q_y)] + 4I_2 \cos(q_x)\cos(q_y) + 2I_3[\cos(2q_x) + \cos(2q_y)]$$
$$+ 4I_5 \cos(2q_x)\cos(2q_y) + 2I_7[\cos(3q_x) + \cos(3q_y)] \qquad (S13)$$

Our objective is to determine if this magnetic interaction becomes the pairing interaction in the heavy fermion superconducting state. Since the STM experiments from which we extracted the dispersions were performed at temperatures below $T_c$, this approach [solving Eq. S2b for $I(\mathbf{r})$ using the normal state Greens functions of Eq. S3b] neglects the feedback effect of superconductivity on the magnetic interaction, $I(\mathbf{q})$. While the inclusion of this feedback effect in the calculation of $I(\mathbf{q})$ and of the superconducting gaps, $\Delta_{\mathbf{k}}^{\alpha,\beta}$, is computationally very demanding, its physical consequences are minor, as discussed in Section S2.

## 2. Heavy Fermions in the Superconducting State of CeCoIn₅

### 2.1 Predicting Superconducting Energy Gaps and $T_c$

Pursuing the idea that it is the magnetic $f$-electron interaction potential Eq. S13 that specifically gives rise to pairing in the CeCoIn₅ SC state, we next predict the superconducting energy gaps and the value of the critical temperature for this material. Such a quantitatively realistic model of the mechanism of superconductivity valid for a specific heavy fermion compound has only now become possible because of accurate determination of the heavy fermion band structure (10) in CeCoIn₅.

While the introduction of the Hubbard-Stratonovich field $t_f(\mathbf{r}, \mathbf{r}')$ leads to an effective decoupling of the magnetic interaction term in the particle-hole channel, the decoupling of the same interaction term in the particle-particle channel leads, in theory, to the emergence of superconductivity. Assuming that the superconducting pairing interaction arises from the spin-flip component of the $f$-electron magnetic interaction $\sum_{\mathbf{r},\mathbf{r}'} I_{\mathbf{r},\mathbf{r}'} \mathbf{S}_\mathbf{r} \cdot \mathbf{S}_{\mathbf{r}'}$ in Eq. S1 given by

$$H_{sf} = \frac{1}{2N} \sum_{\mathbf{k},\mathbf{p},\mathbf{q}} I(\mathbf{q}) \, f^\dagger_{\mathbf{k}+\mathbf{q},\uparrow} f_{\mathbf{k},\downarrow} f^\dagger_{\mathbf{p}-\mathbf{q},\downarrow} f_{\mathbf{p},\uparrow} \tag{S14}$$

we apply the unitary transformation, Eq. S6 to $H_{sf}$ and then decouple $H_{sf}$ in the particle-particle channel. Here, we neglect superconducting pairing terms of the form $\langle \alpha^\dagger_{\mathbf{k},\uparrow} \beta^\dagger_{-\mathbf{k},\downarrow} \rangle$ which are strongly suppressed due to the momentum mismatch of the $\alpha$- and $\beta$-band Fermi surfaces (see Fig. 2 of the main text). The resulting superconducting pairing Hamiltonian then takes the form

$$H_{SC} = -{\sum_\mathbf{p}}' \left( \Delta^\alpha_\mathbf{k} \alpha_{\mathbf{k},\downarrow} \alpha_{-\mathbf{k},\uparrow} + \Delta^\beta_\mathbf{k} \beta_{\mathbf{k},\downarrow} \beta_{-\mathbf{k},\uparrow} + h.c. \right) \tag{S15}$$

where the primed sum is restricted to all states within the Debye energy of the Fermi energy, i.e., to those states with

$$\left| E^{\alpha,\beta}_k \right| \leq \omega_D$$

The superconducting gaps $\Delta_{\mathbf{k}}^{\alpha,\beta}$ are determined via the gap-equations

$$\Delta_{\mathbf{k}}^{\alpha} = -\frac{x_{\mathbf{k}}^2}{N}\sideset{}{'}\sum_{\mathbf{p}} V_{SC}(\mathbf{p}-\mathbf{k})\left(x_{\mathbf{p}}^2\langle\alpha_{\mathbf{p},\uparrow}^{\dagger}\alpha_{-\mathbf{p},\downarrow}^{\dagger}\rangle + w_{\mathbf{p}}^2\langle\beta_{\mathbf{p},\uparrow}^{\dagger}\beta_{-\mathbf{p},\downarrow}^{\dagger}\rangle\right) \quad \text{(S16a)}$$

$$\Delta_{\mathbf{k}}^{\beta} = -\frac{w_{\mathbf{k}}^2}{N}\sideset{}{'}\sum_{\mathbf{p}} V_{SC}(\mathbf{p}-\mathbf{k})\left(x_{\mathbf{p}}^2\langle\alpha_{\mathbf{p},\uparrow}^{\dagger}\alpha_{-\mathbf{p},\downarrow}^{\dagger}\rangle + w_{\mathbf{p}}^2\langle\beta_{\mathbf{p},\uparrow}^{\dagger}\beta_{-\mathbf{p},\downarrow}^{\dagger}\rangle\right) \quad \text{(S16b)}$$

where $N$ is the number of sites in the system, and $V_{SC}(\mathbf{q}) = -I(\mathbf{q})/2$ is the effective superconducting pairing potential, as shown in Fig. 3A of the main text.

These are the essential equations from which the predictions about the superconducting electronic structure and Cooper pairing mechanism are derived, and are summarized as Eq. 2 of the main text. The total Hamiltonian $H_{SC}^{MF} = H_{MF} + H_{SC}$ (with $H_{MF}$ in Eq. S5) can then be diagonalized using separate Bogoliubov transformations for the $\alpha$- and $\beta$-bands given by

$$\begin{aligned}
\alpha_{\mathbf{k},\uparrow} &= u_{\mathbf{k}}^{\alpha} a_{\mathbf{k}} + v_{\mathbf{k}}^{\alpha} b_{\mathbf{k}}^{\dagger} \\
\alpha_{-\mathbf{k},\downarrow} &= v_{\mathbf{k}}^{\alpha} a_{\mathbf{k}}^{\dagger} - u_{\mathbf{k}}^{\alpha} b_{\mathbf{k}} \\
\beta_{\mathbf{k},\uparrow} &= u_{\mathbf{k}}^{\beta} d_{\mathbf{k}} + v_{\mathbf{k}}^{\beta} g_{\mathbf{k}}^{\dagger} \\
\beta_{-\mathbf{k},\downarrow} &= v_{\mathbf{k}}^{\beta} d_{\mathbf{k}}^{\dagger} - u_{\mathbf{k}}^{\beta} g_{\mathbf{k}}
\end{aligned} \quad \text{(S17)}$$

yielding

$$H_{SC}^{MF} = \sideset{}{'}\sum_{\mathbf{p}} \left[\Omega_{\mathbf{k}}^{\alpha}(a_{\mathbf{k}}^{\dagger}a_{\mathbf{k}} + b_{\mathbf{k}}^{\dagger}b_{\mathbf{k}}) + \Omega_{\mathbf{k}}^{\beta}(d_{\mathbf{k}}^{\dagger}d_{\mathbf{k}} + g_{\mathbf{k}}^{\dagger}g_{\mathbf{k}})\right] \quad \text{(S18)}$$

with

$$\Omega_{\mathbf{k}}^{\alpha,\beta} = \sqrt{\left(E_{\mathbf{k}}^{\alpha,\beta}\right)^2 + \left(\Delta_{\mathbf{k}}^{\alpha,\beta}\right)^2} \quad \text{(S19)}$$

Applying the same unitary transformation to the gap equation yields

$$\Delta_{\mathbf{k}}^{\alpha} = -\frac{x_{\mathbf{k}}^2}{N} {\sum_{\mathbf{p}}}' V_{SC}(\mathbf{p}-\mathbf{k}) \left[ x_{\mathbf{p}}^2 \frac{\Delta_{\mathbf{p}}^{\alpha}}{2\Omega_{\mathbf{p}}^{\alpha}} \tanh\left(\frac{\Omega_{\mathbf{p}}^{\alpha}}{2k_B T}\right) + w_{\mathbf{p}}^2 \frac{\Delta_{\mathbf{p}}^{\beta}}{2\Omega_{\mathbf{p}}^{\beta}} \tanh\left(\frac{\Omega_{\mathbf{p}}^{\beta}}{2k_B T}\right) \right] \quad \text{(S20a)}$$

$$\Delta_{\mathbf{k}}^{\beta} = -\frac{w_{\mathbf{k}}^2}{N} {\sum_{\mathbf{p}}}' V_{SC}(\mathbf{p}-\mathbf{k}) \left[ x_{\mathbf{p}}^2 \frac{\Delta_{\mathbf{p}}^{\alpha}}{2\Omega_{\mathbf{p}}^{\alpha}} \tanh\left(\frac{\Omega_{\mathbf{p}}^{\alpha}}{2k_B T}\right) + w_{\mathbf{p}}^2 \frac{\Delta_{\mathbf{p}}^{\beta}}{2\Omega_{\mathbf{p}}^{\beta}} \tanh\left(\frac{\Omega_{\mathbf{p}}^{\beta}}{2k_B T}\right) \right] \quad \text{(S20b)}$$

Notice that the coherence factors $x_{\mathbf{k}}^2, w_{\mathbf{k}}^2$ on the r.h.s. of Eq. S20 arise from the projection of the $f$-electron pairing interaction (Eq. S14) onto the $\alpha$- and $\beta$-bands. Below, we solve Eq. S20a and Eq. S20b to obtain $\Delta_{\mathbf{k}}^{\alpha,\beta}$ for all momentum states within $\omega_D$ the Fermi energy. To account for finite lifetime effects of the $c$- and $f$-electron states (for example, arising from defects, interactions with phonons, etc.), we write the gap equations Eqs. S16a and S16b in an alternative but equivalent form given by

$$\Delta_{\mathbf{k}}^{\alpha} = -\frac{x_{\mathbf{k}}^2}{N} {\sum_{\mathbf{p}}}' V_{SC}(\mathbf{p}-\mathbf{k}) \left\{ -x_{\mathbf{p}}^2 \frac{\Delta_{\mathbf{p}}^{\alpha}}{2\Omega_{\mathbf{p}}^{\alpha}} \int_{-\infty}^{\infty} \frac{d\omega}{\pi} \text{Im} G_a(\mathbf{p},\omega) \tanh\left(\frac{\omega}{2k_B T}\right) \right.$$
$$\left. - w_{\mathbf{p}}^2 \frac{\Delta_{\mathbf{p}}^{\beta}}{2\Omega_{\mathbf{p}}^{\beta}} \int_{-\infty}^{\infty} \frac{d\omega}{\pi} \text{Im} G_d(\mathbf{p},\omega) \tanh\left(\frac{\omega}{2k_B T}\right) \right\} \quad \text{(S21a)}$$

$$\Delta_{\mathbf{k}}^{\beta} = -\frac{w_{\mathbf{k}}^2}{N} {\sum_{\mathbf{p}}}' V_{SC}(\mathbf{p}-\mathbf{k}) \left\{ -x_{\mathbf{p}}^2 \frac{\Delta_{\mathbf{p}}^{\alpha}}{2\Omega_{\mathbf{p}}^{\alpha}} \int_{-\infty}^{\infty} \frac{d\omega}{\pi} \text{Im} G_a(\mathbf{p},\omega) \tanh\left(\frac{\omega}{2k_B T}\right) \right.$$
$$\left. + w_{\mathbf{p}}^2 \frac{\Delta_{\mathbf{p}}^{\beta}}{2\Omega_{\mathbf{p}}^{\beta}} \int_{-\infty}^{\infty} \frac{d\omega}{\pi} \text{Im} G_d(\mathbf{p},\omega) \tanh\left(\frac{\omega}{2k_B T}\right) \right\} \quad \text{(S21b)}$$

where

$$G_a(\mathbf{p},\omega) = G_b(\mathbf{p},\omega) = \frac{1}{\omega - \Omega_{\mathbf{p}}^{\alpha} + i\Gamma} \quad \text{(S22a)}$$

$$G_d(\mathbf{p},\omega) = G_g(\mathbf{p},\omega) = \frac{1}{\omega - \Omega_{\mathbf{p}}^{\beta} + i\Gamma} \quad \text{(S22b)}$$

with $\Gamma$ being the inverse lifetime of the quasi-particle states. We here assume that the microscopic mechanism giving rise to a non-zero $\Gamma$ does not lead to a renormalization of the pairing vertex. This is the case, for example, when $\Gamma$ arises from scattering of non-magnetic defects with a momentum independent scattering potential, of from scattering of magnetic defects. In all other cases, a renormalization of the pairing vertex is possible, with the strength of this renormalization (and its effect on $T_c$) dependent on the microscopic form of the scattering potential (11). Finally, we note that in the limit $\Gamma \to 0$ Eq. S21a and Eq. S21b become the standard gap equations Eq. S20a and Eq. S20b. In what follows, we assume that any mechanism leading to decoherence and thus a non-zero $\Gamma$ is suppressed by the opening of the SC gap such that for $T = 0$ we set $\Gamma = 0^+$.

Our solutions of the gap equations, Eq. S20a and Eq. S20b, predict that the $\alpha$ and $\beta$ bands of CeCoIn$_5$ possess superconducting gaps $\Delta_{\mathbf{k}}^{\alpha,\beta}$ of nodal $d_{x^2-y^2}$-symmetry as shown in Fig. 3B,C of the main text and that the maximum gap value occurs on the $\alpha_1$-Fermi surface, with $\Delta_{\mathbf{k}}^{\alpha,\beta}$ being well approximated by

$$\Delta_{\mathbf{k}}^{\alpha} = \frac{\Delta_0^{\alpha}}{2}\{[\cos(k_x) - \cos(k_y)] + \alpha_1[\cos(2k_x) - \cos(2k_y)] + \alpha_2[\cos(3k_x) - \cos(3k_y)]\}$$

(S23a)

$$\Delta_{\mathbf{k}}^{\beta} = \frac{\Delta_0^{\beta}}{2}[\cos(k_x) - \cos(k_y)]^3$$

(S23b)

Here, $\Delta_0^{\alpha} = 0.492$ meV, $\alpha_1 = -0.607$, $\alpha_2 = -0.082$, $\Delta_0^{\beta} = -1.040$ meV represent the quantitative predictions for $\Delta_{\mathbf{k}}^{\alpha,\beta}$ obtained for $\omega_D = 0.66$ meV, yielding a maximum superconducting energy gap of $\Delta_{max} \leq 600$ $\mu$eV. Note that a value of $\omega_D = 0.66$ meV is consistent with the energy scale of the spin fluctuation spectrum (SOM Section S4).

The emergence of a superconducting gap with $d_{x^2-y^2}$-symmetry, and in particular its nodal structure, can be simply understood by considering the structure of the pairing interaction, $V_{SC}(\mathbf{r}) = -I(\mathbf{r})/2$, in real space. For example, an antiferromagnetic interaction between nearest neighbors ($I_1 > 0$) and a ferromagnetic interaction between next-nearest neighbors, ($I_2 < 0$), translates into an attractive pairing potential along the bond direction ($V_{SC,1} < 0$), and a repulsive pairing potential ($V_{SC,2} > 0$) along the diagonal direction of the underlying lattice thus yielding a nodal structure of the superconducting gaps. The emergence of the $d_{x^2-y^2}$-symmetry can also be understood in momentum space: the large repulsive pairing potential near $\mathbf{Q} = (1,1)\pi/a_0$ (see arrow in Fig. 3A) requires, as follows from the BCS gap equation, Eq. S20, that the superconducting gap changes sign between Fermi surface points connected by $\mathbf{Q}$, i.e., for $\mathbf{p} - \mathbf{k} = \mathbf{Q}$ as shown in Fig. 3B.

It is worth noting that the maximum superconducting gap on the $\beta$-Fermi surface, $\left|\Delta_{max}^{\beta}\right| \approx 95$ μeV is significantly smaller than the maximum superconducting gap $|\Delta_{max}| \approx 600$ μeV which occurs on the $\alpha_1$ Fermi surface. The reason for this is twofold. First, the states on the $\beta$-Fermi surface consist primarily of $c$-electron states, which reduces their effective coupling (via the coherence factors $w_k^2$ in the BCS gap equation Eq. S20b) to the superconducting pairing potential, $V_{SC}$, which originates from the magnetic interaction in the $f$-electron band. Second, the magnitude of the pairing potential, $V_{SC}$, is smaller for small momenta, and the wave-vectors connecting states on the $\beta$-Fermi surface are smaller than those connecting states on the $\alpha$-Fermi surface. Thus, the magnitude of the pairing potential, $V_{SC}$, is smaller for pairing within the $\beta$-band, leading to a value of $\Delta_{max}^{\beta}$ that is significantly smaller than that of $\Delta_{max}^{\alpha}$. We note that there is no direct signature of a non-zero gap on the $\beta$-band in our tunneling experiments since a value of $\left|\Delta_{max}^{\beta}\right| \approx 95$ μeV is to close to the experimental detection limit (12) at 250 mK. Moreover, the phase shift between the superconducting gaps on the $\alpha$- and $\beta$-Fermi surfaces originates from the fact that the pairing potential for momenta $\mathbf{q}$ connecting points on the $\alpha$- and $\beta$-Fermi surfaces is predominantly repulsive. The solution of the BCS gap equation Eq. S20 thus requires that the superconducting gaps at these points possess different signs.

With $\omega_D = 0.66$ meV fixed, we solve the gap equation for $\Gamma = 0^+$ with no further adjustable parameters, obtaining a critical temperature of $T_c = 2.96$ K. With increasing $\Gamma$ (corresponding to a decreasing lifetime of the quasi-particles) the critical temperature is suppressed as expected due to the ensuing decoherence. For $\Gamma = 0.05$ meV (corresponding to the experimentally observed (13) mean-free path of approximately $l = 81$ nm) we obtain from Eq. S21a and Eq. S21b a critical temperature of $T_c = 2.55$K in good agreement with the experimentally observed critical temperature $T_c = 2.3$ K. Thus, within our approach, we obtain

$$\frac{2\Delta_0}{k_B T_c} = 5.46 \tag{S24}$$

by using $\Delta_0 = \Delta_{max} = 0.6$ meV and $T_c = 2.55$K. We note that the ratio $2\Delta_0/k_B T_c$ possesses only a weak dependence on $\omega_D$ and can thus be considered a general result of our theory.

This weak dependence of $2\Delta_0/k_B T_c$ on $\omega_D$ (which is in contrast to the conventional BCS result) arises from the curvature of the energy bands near the Fermi surface and the resulting energy dependence of the density of states.

We note that the above ratio of $2\Delta_0/k_B T_c$ (or the experimentally determined ratio $2\Delta_0/k_B T_c \approx 6.05$) by itself is not sufficient to determine whether CeCoIn$_5$ is a weak, intermediate or strong coupling superconductor. The reason is that in multi-band systems with multiple superconducting gaps, the ratio $2\Delta_0/k_B T_c$ with $\Delta_0$ being the maximum magnitude of the superconducting gap in all bands, can significantly exceed the result for a weak-coupling, single band superconductor (14,15). Here, we recall that for a weak-coupling one-band superconductor with $d_{x^2-y^2}$-symmetry, one finds $2\Delta_0/k_B T_c \approx 4.3$, Ref.(16), in contrast to the BCS result of $2\Delta_0/k_B T_c \approx 3.53$ for a single-band s-wave superconductor. Even taking this ratio of 4.3 as the lower bound for CeCoIn$_5$ would imply that it is at the most a moderately coupled superconductor.

Furthermore, we note that while dHvA experiments have observed both 2D and 3D bands in CeCoIn$_5$, the combination of theoretical results presented in the main text and the subsequent sections and their good agreement with experimental findings clearly suggests that the bands relevant for the emergence of superconductivity are 2D in nature.

In order to study the feedback effect of superconductivity on the effective pairing interaction $V_{SC}(\mathbf{q}) = -I(\mathbf{q})/2$, we solve Eq. S2b in the superconducting state, using the superconducting Green's functions given in Eq. S29. However, since the latter depend on the superconducting gap themselves, it is now necessary to simultaneously solve both the BCS gap equations, Eq. S20, and Eq. S2b self-consistently. This calculation has to be repeated for every temperature below $T_c$ and is thus computationally very demanding. However, executing this approach for $T = 0$ shows that the results for the superconducting gaps are basically identical to those obtained without including a feedback of superconductivity on $V_{SC}(\mathbf{q})$ if the Debye energy is slightly decreased. Thus, to reduce the computational demands of our approach, we have neglected the feedback effect in the calculation of the superconducting gaps.

Finally, we note that while both the bandwidth $W_f$ of $\varepsilon_\mathbf{k}^f$ and the superconducting gaps, $\Delta_\mathbf{k}^{\alpha,\beta}$, arise from the magnetic interactions, $I(\mathbf{r})$, the respective energy scales, $W_f \approx 11$ meV and $\Delta_{max} \approx 0.6$ meV are quite different. The reason for this lies in the functional relationships between $I(\mathbf{r})$ and $W_f$ on one hand, and $\Delta_\mathbf{k}^{\alpha,\beta}$ on the other hand. To exemplify this, consider the simple case where only $I_1 \neq 0$, while $I_i = 0$ for all other terms. In this case, one has $W_f = 8t_{f1}$ and for vanishing hybridization, $s_\mathbf{k} \to 0$, one obtains $t_{f1} = 2I_1/\pi^2$ from Eq.S2b. On the other hand, we find that the scaling between the maximum superconducting gap and the maximum value, $I_{max}$ of $I(\mathbf{q})$ at $\mathbf{q} = (1,1)\pi/a_0$ is approximately given by

$$\Delta_{max} \sim \exp\left(-\frac{1}{N_0 I_{max}}\right)$$

thus explaining the difference in the scales of $W_f$ and $\Delta_{max}$.

## 2.2 Bogoliubov quasi-particle interference (BQPI) in the superconducting state of CeCoIn$_5$

After having obtained the momentum dependence of the superconducting gaps below $T_c$, we can now predict the form of the quasi-particle interference spectrum [the so-called *Bogoliubov quasi-particle interference* (BQPI) spectrum] in the superconducting state. In the Born approximation, the BQPI spectrum $g(\mathbf{q}, E) = |\bar{g}(\mathbf{q}, E)|$ for each of the spin projections is given by

$$\bar{g}(\mathbf{q}, E) \equiv \frac{dI(\mathbf{q}, E)}{dV} = \frac{\pi e}{\hbar} N_t \sum_{i,j=1}^{2} \left[\hat{t}\widehat{N}(\mathbf{q}, E)\hat{t}\right]_{ij} \qquad (S25a)$$

$$\widehat{N}(\mathbf{q}, E) = -\frac{1}{\pi} \text{Im}\left[\frac{1}{N}\sum_\mathbf{k} \widehat{G}(\mathbf{k}, E)\widehat{U}\widehat{G}(\mathbf{k} + \mathbf{q}, E)\right] \qquad (S25b)$$

where now

$$\widehat{U} = \begin{pmatrix} U_{cc} & U_{cf} & 0 & 0 \\ U_{fc} & U_{ff} & 0 & 0 \\ 0 & 0 & -U_{cc} & -U_{cf} \\ 0 & 0 & -U_{fc} & -U_{ff} \end{pmatrix}, \qquad (S26a)$$

$$\hat{t} = \begin{pmatrix} -t_c & 0 & 0 & 0 \\ 0 & -t_f & 0 & 0 \\ 0 & 0 & t_c & 0 \\ 0 & 0 & 0 & t_f \end{pmatrix}, \qquad (S26b)$$

and

$$\hat{G}(\mathbf{k}, E) = \begin{pmatrix} G_{cc}(\mathbf{k}, \sigma, E) & G_{cf}(\mathbf{k}, \sigma, E) & F_{cc}(\mathbf{k}, E) & F_{cf}(\mathbf{k}, E) \\ G_{fc}(\mathbf{k}, \sigma, E) & G_{ff}(\mathbf{k}, \sigma, E) & F_{fc}(\mathbf{k}, E) & F_{ff}(\mathbf{k}, E) \\ F_{cc}(\mathbf{k}, E) & F_{cf}(\mathbf{k}, E) & -G_{cc}(\mathbf{k}, \sigma, -E) & -G_{cf}(\mathbf{k}, \sigma, -E) \\ F_{fc}(\mathbf{k}, E) & F_{ff}(\mathbf{k}, E) & -G_{fc}(\mathbf{k}, \sigma, -E) & -G_{ff}(\mathbf{k}, \sigma, -E) \end{pmatrix} \quad (S27)$$

Here, the normal Green's functions ($\gamma, \zeta = c, f$)

$$G_{\gamma\zeta}(\mathbf{r}', \mathbf{r}, \sigma, \tau) = -\langle T_\tau \gamma_{\mathbf{r}', \sigma}(\tau) \zeta_{\mathbf{r}, \sigma}^\dagger(0) \rangle, \quad (S28a)$$

and anomalous Green's functions

$$F_{\gamma\zeta}(\mathbf{r}', \mathbf{r}, \tau) = -\langle T_\tau \gamma_{\mathbf{r}', \uparrow}^\dagger(\tau) \zeta_{\mathbf{r}, \downarrow}^\dagger(0) \rangle, \quad (S28b)$$

reflecting both the hybridization between the $c$- and $f$-electron bands as well as the emergence of the superconducting order parameter in the $\alpha$- and $\beta$-bands, are given by

$$G_{cc}(\mathbf{k}, \sigma, \omega) = w_\mathbf{k}^2 \frac{\omega + i\Gamma + E_\mathbf{k}^\alpha}{(\omega + i\Gamma)^2 - (\Omega_\mathbf{k}^\alpha)^2} + x_k^2 \frac{\omega + i\Gamma + E_\mathbf{k}^\beta}{(\omega + i\Gamma)^2 - (\Omega_\mathbf{k}^\beta)^2} \quad (S29a)$$

$$G_{cf}(\mathbf{k}, \sigma, \omega) = w_\mathbf{k} x_\mathbf{k} \left[ \frac{\omega + i\Gamma + E_\mathbf{k}^\alpha}{(\omega + i\Gamma)^2 - (\Omega_\mathbf{k}^\alpha)^2} - \frac{\omega + i\Gamma + E_\mathbf{k}^\beta}{(\omega + i\Gamma)^2 - (\Omega_\mathbf{k}^\beta)^2} \right] \quad (S29b)$$

$$G_{ff}(\mathbf{k}, \sigma, \omega) = x_\mathbf{k}^2 \frac{\omega + i\Gamma + E_\mathbf{k}^\alpha}{(\omega + i\Gamma)^2 - (\Omega_\mathbf{k}^\alpha)^2} + w_k^2 \frac{\omega + i\Gamma + E_\mathbf{k}^\beta}{(\omega + i\Gamma)^2 - (\Omega_\mathbf{k}^\beta)^2} \quad (S29c)$$

$$F_{cc}(\mathbf{k}, \omega) = w_\mathbf{k}^2 \frac{\Delta_\mathbf{k}^\alpha}{(\omega + i\Gamma)^2 - (\Omega_\mathbf{k}^\alpha)^2} + x_\mathbf{k}^2 \frac{\Delta_\mathbf{k}^\beta}{(\omega + i\Gamma)^2 - (\Omega_\mathbf{k}^\beta)^2} \quad (S29d)$$

$$F_{cf}(\mathbf{k}, \omega) = w_\mathbf{k} x_\mathbf{k} \left[ \frac{\Delta_\mathbf{k}^\alpha}{(\omega + i\Gamma)^2 - (\Omega_\mathbf{k}^\alpha)^2} - \frac{\Delta_\mathbf{k}^\beta}{(\omega + i\Gamma)^2 - (\Omega_\mathbf{k}^\beta)^2} \right] \quad (S29e)$$

$$F_{ff}(\mathbf{k}, \omega) = x_\mathbf{k}^2 \frac{\Delta_\mathbf{k}^\alpha}{(\omega + i\Gamma)^2 - (\Omega_\mathbf{k}^\alpha)^2} + w_\mathbf{k}^2 \frac{\Delta_\mathbf{k}^\beta}{(\omega + i\Gamma)^2 - (\Omega_\mathbf{k}^\beta)^2} \quad (S29f)$$

where $\Gamma^{-1}$ is the lifetime of the c- and f-electron states. Note that in the momentum sum in Eq. S25b, $\Delta_{\mathbf{k}}^{\alpha,\beta} \neq 0$ only for those momentum states within the Debye energy of the Fermi energy.

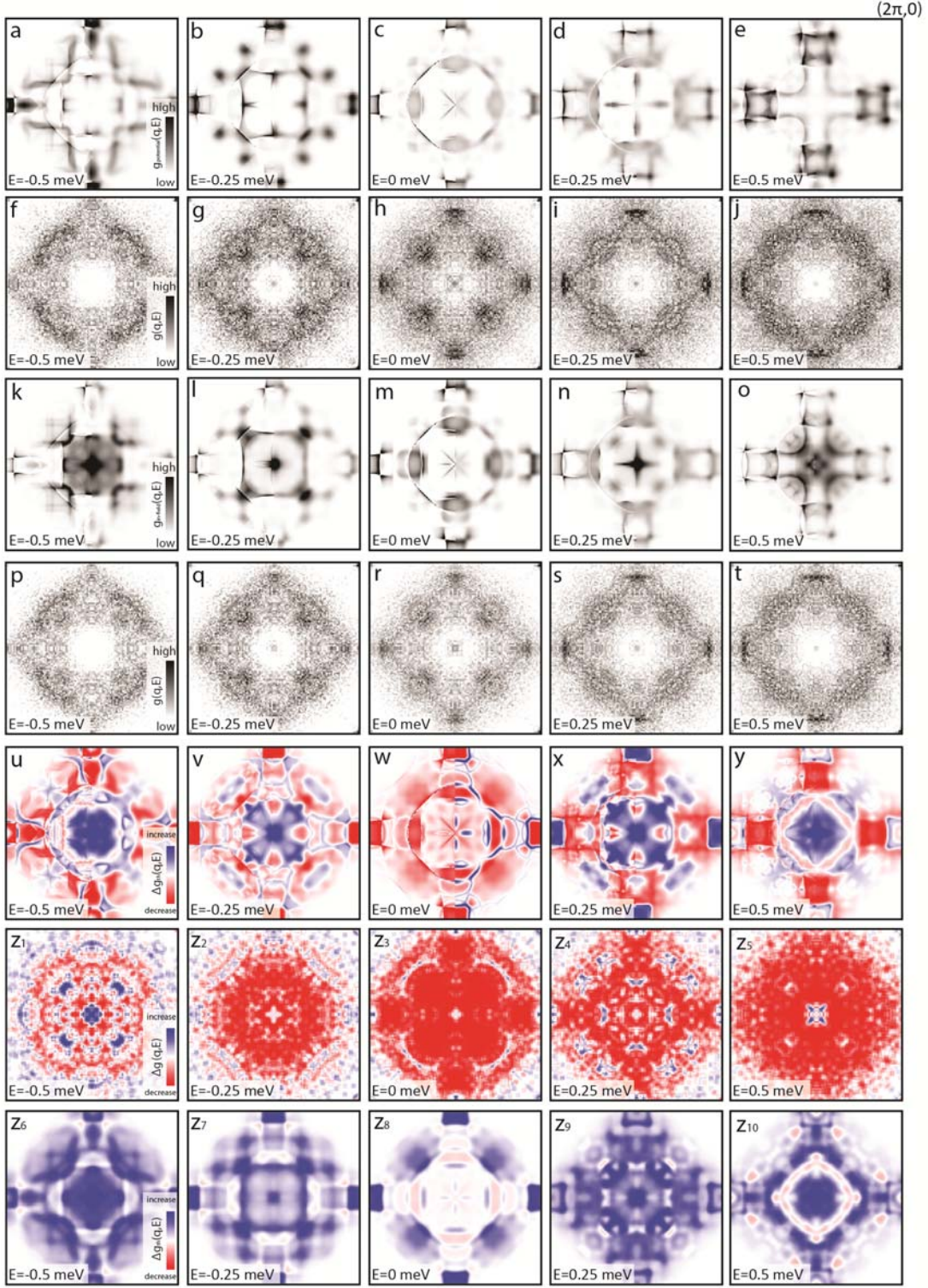

**Fig. S2.** (a-e) Theoretical $g(\mathbf{q}, E)$ for pure potential scattering (no magnetic field, $B = 0$), (f-j) experimental $g(\mathbf{q}, E)$ for $B = 0$. (k-o) Theoretical $g(\mathbf{q}, E)$ (in-field, $\mathbf{B} \neq 0$) for a combination of magnetic and potential scattering, as described in the text, (p-t) experimental $g(\mathbf{q}, E)$ for $B \neq 0$. For the in-field simulations, $M_{cc} = -1.7 U_{cc}$ was used giving the best correspondence to the experimental data. Modified theoretical (u-y) and experimental (z1-z5) PQPI intensity $\Delta g(\mathbf{q}, E, B)$, as described in the text and Fig. S3. Modified theoretical (z6-z10) PQPI intensity $\Delta g(\mathbf{q}, E, B)$ for a nodal $s$-wave superconductor as described in the text. All theoretical $g(\mathbf{q}, E)$ have been plotted in a repeated zone scheme using a structure factor of the form $S(\mathbf{q}) = \frac{1}{2}\sqrt{\left[1 + \cos\left(\frac{q_x}{2}\right)\right]\left[1 + \cos\left(\frac{q_y}{2}\right)\right]}$ to suppress high-q scattering generally observed in experiment. The right half of each panel is low-pass filtered to mimic the finite experimental resolution.

The theoretical BQPI data for a series of energies inside the SC gap are shown in Fig. S2(a)-(e), while the experimental BQPI results are shown in Fig. S2(f)-(j). The good agreement between the two data sets further supports the theoretically predicted form of the superconducting gaps.

## 2.3 Phase-sensitive Quasi-Particle Interference (PQPI)

Bogoliubov quasi-particle interference scattering is a powerful tool to extract the gap symmetry of a superconductor, but is not sensitive to the phase of the order parameter. Phase-sensitive quasi-particle interference scattering on the other hand can in principle give information about the phase of the order parameter, and thereby enable one to distinguish between a sign-changing $d_{x^2-y^2}$ and a nodal $s$-wave superconducting gap (17,18). The underlying idea of PQPI is that there is a different contribution from potential and magnetic scattering at zero external magnetic field and finite magnetic field in a quasi-particle interference scattering experiment. Since the magnetic scattering component is sensitive to a sign changing of the order parameter in a different way than the potential scattering component, comparison of data sets taken in a field and in zero field should enable one to extract the sign of the order parameter. Here we will first describe our theoretical expectations (section 2.3.1) based on the BQPI discussed in the previous section, and then compare these to our experimentally measured PQPI spectra (Section 2.3.2).

### 2.3.1 Theoretical simulation of PQPI

In the presence of a magnetic field, the QPI spectrum is controlled by a superposition of a term due to non-magnetic scattering, $\bar{g}_{pot}(\mathbf{q}, E)$, which is also present for $B = 0$, and a magnetic-field induced contribution, $\bar{g}_{mag}(\mathbf{q}, E, B)$. The latter is computed from Eq. S25 by using a magnetic scattering matrix

$$\hat{U}_M = \begin{pmatrix} M_{cc} & M_{cf} & 0 & 0 \\ M_{fc} & M_{ff} & 0 & 0 \\ 0 & 0 & M_{cc} & M_{cf} \\ 0 & 0 & M_{fc} & M_{ff} \end{pmatrix} \tag{S30}$$

where $M_{cc}$ and $M_{ff} = M_{ff}^{(0)} z_0^2$ are the magnetic scattering potential for intraband scattering in the $c$- and $f$-electron bands, respectively, while $M_{fc} = M_{cf} = M_{cf}^{(0)} z_0$ is the magnetic scattering potential for interband scattering between the $c$- and $f$-electron bands.

Thus, one has
$$\bar{g}(\mathbf{q}, E, B \neq 0) = \bar{g}_{pot}(\mathbf{q}, E) + \bar{g}_{mag}(\mathbf{q}, E, B) \tag{S31a}$$
$$\bar{g}(\mathbf{q}, E, B = 0) = \bar{g}_{pot}(\mathbf{q}, E) \tag{S31b}$$

While the precise magnetic field dependence of $\bar{g}_{mag}(\mathbf{q}, E, B)$ is currently unknown, a good theoretical description of the experimental BQPI data in a magnetic field of $B = 3T$ can be achieved by setting for simplicity $M_{cf}/M_{cc} = U_{cf}/U_{cc}$ and $M_{ff}/M_{cc} = U_{ff}/U_{cc}$, and by choosing $M_{cc} \approx -1.7 U_{cc}$. A comparison of the theoretical BQPI layers in a magnetic field of $B = 3T$, i.e., $g(\mathbf{q}, E, B \neq 0) = |\bar{g}(\mathbf{q}, E, B \neq 0)|$ of Eq. S31a, shown in Figs. S2(k)- (o) with the experimental results shown in Figs. S2(p) – (t) demonstrates good agreement between the theoretical and experimental results.

We next consider predictions for the phase-sensitive QPI spectrum of CeCoIn5, defined as

$$\Delta g(\mathbf{q}, E, B) = g(\mathbf{q}, E, B) - g(\mathbf{q}, E, 0) = |\bar{g}(\mathbf{q}, E, B)| - |\bar{g}(\mathbf{q}, E, 0)| \tag{S32}$$

To develop a better understanding for the microscopic origin of $\Delta g(\mathbf{q}, E, B)$, we note that both the BQPI spectra for magnetic and non-magnetic scattering as derived from Eq. S26 can be written as a sum of two contributions: one, $\bar{g}_G(\mathbf{q}, E)$, involving a combination of normal Green's functions [Eq. S29a – Eq. S29c] only, and one, $\bar{g}_F(\mathbf{q}, E)$, involving a combination of anomalous Green's functions [Eq. S29d – Eq. S29f] only. As we demonstrate below, $\bar{g}_F(\mathbf{q}, \omega)$ is a sensitive probe for the phase change of the superconducting order parameter, while $\bar{g}_G(\mathbf{q}, E)$ is not. We note that $\bar{g}_F(\mathbf{q}, \omega)$ enters the BQPI spectra for magnetic and non-magnetic scattering with opposite sign, such that

$$\bar{g}_{pot}(\mathbf{q}, E) = U_{cc}[\bar{g}_G(\mathbf{q}, E) - \bar{g}_F(\mathbf{q}, E)] \tag{S33a}$$

$$\bar{g}_{mag}(\mathbf{q}, E, B) = M_{cc}[\bar{g}_G(\mathbf{q}, E) + \bar{g}_F(\mathbf{q}, E)] \tag{S33b}$$

where we have used $M_{cf}/M_{cc} = U_{cf}/U_{cc}$ and $M_{ff}/M_{cc} = U_{ff}/U_{cc}$. We thus obtain

$$\begin{aligned}\Delta g(\mathbf{q}, E, B) &= |\bar{g}(\mathbf{q}, E, B)| - |\bar{g}(\mathbf{q}, E, 0)| \\ &= |U_{cc}[\bar{g}_G(\mathbf{q}, E) - \bar{g}_F(\mathbf{q}, E)] + M_{cc}[\bar{g}_G(\mathbf{q}, E) + \bar{g}_F(\mathbf{q}, E)]| \\ &\quad - |U_{cc}[\bar{g}_G(\mathbf{q}, E) - \bar{g}_F(\mathbf{q}, E)]|\end{aligned} \tag{S34}$$

In the above expression, only $\bar{g}_F(\mathbf{q}, E)$ is sensitive to the phase of the superconducting order parameter, while $\bar{g}_G(\mathbf{q}, E)$ is not (see below). The general forms of $\bar{g}_{G,F}(\mathbf{q}, E)$ are rather complex (as follows from Eq. S25 – Eq. S29), since due to the two-band electronic structure of CeCoIn$_5$, they contain contributions from intra- and inter-band scattering. However, to exemplify the phase-sensitivity of $\bar{g}_F(\mathbf{q}, E)$ it is sufficient to consider scattering processes involving only one of the bands. We therefore choose an energy $E$, and two scattering vectors $\mathbf{q}_{1,2}$ for which the main contribution to $\bar{g}_F(\mathbf{q}, E)$ comes from scattering processes involving momentum points near the $\alpha$-FS, in which case

$$\bar{g}_F(\mathbf{q}, E) \approx \bar{g}_F^{\alpha\alpha}(\mathbf{q}, E) = \mathrm{Im}\left[\frac{1}{N}\sum_{\mathbf{k}} B(\mathbf{k}, \mathbf{k+q}) \frac{\Delta_{\mathbf{k}}^{\alpha}}{(E+i\Gamma)^2 - (\Omega_{\mathbf{k}}^{\alpha})^2} \frac{\Delta_{\mathbf{k+q}}^{\alpha}}{(E+i\Gamma)^2 - (\Omega_{\mathbf{k+q}}^{\alpha})^2}\right] \tag{S35}$$

Here $B(\mathbf{k}, \mathbf{k+q})$ contains terms involving the tunneling amplitudes, $t_c, t_f$, as well as the coherence factors $x_{\mathbf{k}}, w_{\mathbf{k}}$ arising from the hybridization of the light and heavy bands; it therefore does not contain phase sensitive information. In Fig. 4C of the main text we present the equal energy contour (EEC) for $E = -\Omega_{\mathbf{k}}^{\alpha} = -0.5$ meV together wth the scattering processes involving $\mathbf{q}_{1,2}$ that yield the dominant contribution to $\bar{g}_F^{\alpha\alpha}(\mathbf{q}_{1,2}, E)$. Since $\Delta_{\mathbf{k}}^{\alpha}$ possesses a $d_{x^2-y^2}$-wave symmetry, scattering vector $\mathbf{q}_1$ connects momentum points on the EEC where $\Delta_{\mathbf{k}}^{\alpha}$ possesses different phases, while $\mathbf{q}_2$ connects momentum points where $\Delta_{\mathbf{k}}^{\alpha}$ possesses the same phase. As a result, the sign of $\bar{g}_F(\mathbf{q}_1, E) \approx \bar{g}_F^{\alpha\alpha}(\mathbf{q}_1, E)$ is different from that of $\bar{g}_F(\mathbf{q}_2, E)$, reflecting the relative change of phase of $\Delta_{\mathbf{k}}^{\alpha}$ along the Fermi surface. Applying the same argument to $\bar{g}_G(\mathbf{q}, E)$ yields that $\bar{g}_G(\mathbf{q}_1, E)$ possesses the same sign as $\bar{g}_G(\mathbf{q}_2, E)$ and is thus not sensitive to a phase change of $\Delta_{\mathbf{k}}^{\alpha}$.

Based on the above discussion, it is clear that $\Delta g(\mathbf{q}, E, B)$ in Eq. S34 contains in general both phase-sensitive and phase-insensitive contributions, making it experimentally challenging to unambiguously identify a phase change of the superconducting order parameter. However, for $M_{cc} = -2U_{cc}$, one finds that $\bar{g}(\mathbf{q}, E, B)$ and $\bar{g}(\mathbf{q}, E, B = 0)$ in Eq. S34 possess opposite signs (note that $\bar{g}$ is real), and one obtains (with $\zeta = \text{sgn}[\bar{g}(\mathbf{q}, E, B)]$)

$$\begin{aligned}
\Delta g(\mathbf{q}, E, B) &= \zeta\{U_{cc}[\bar{g}_G(\mathbf{q}, E) - \bar{g}_F(\mathbf{q}, E)] + M_{cc}[\bar{g}_G(\mathbf{q}, E) + \bar{g}_F(\mathbf{q}, E)] \\
&\quad + U_{cc}[\bar{g}_G(\mathbf{q}, E) - \bar{g}_F(\mathbf{q}, E)]\} \\
&= \zeta[(2U_{cc} + M_{cc})\bar{g}_G(\mathbf{q}, E) + (M_{cc} - 2U_{cc})\bar{g}_F(\mathbf{q}, E)] \\
&= 2\zeta M_{cc}\bar{g}_F(\mathbf{q}, E)
\end{aligned} \quad (S36)$$

in which case the PQPI spectrum, $\Delta g(\mathbf{q}, E, B)$, is indeed a sensitive probe for the phase of the superconducting gaps.

A comparison of the theoretically predicted, Figs. S2 (u) – (y), and experimentally observed, Figs. S2 (z1) – (z5), PQPI layers for $M_{cc} \approx -1.7U_{cc}$, are in excellent agreement. As this ratio of $M_{cc}/U_{cc}$ is sufficiently close to $-2$, the PQPI spectra directly reflect the phase of the order parameter, which for CeCoIn$_5$ is thus $d_{x^2-y^2}$. This is demonstrated in Fig. 4A and B of the main text where we present the theoretical and experimental $\Delta g(\mathbf{q}, E = -0.5\text{meV}, B)$: $\Delta g(\mathbf{q}_1, E = -0.5\text{meV}, B) < 0$ (red) for the scattering vector $\mathbf{q}_1$ connecting momentum points near the $\alpha$-FS where $\Delta_{\mathbf{k}}^{\alpha}$ possesses different phases, while $\Delta g(\mathbf{q}_2, E = -0.5\text{meV}) > 0$ (blue) for the scattering vector $\mathbf{q}_2$ connecting momentum points near the $\alpha$-FS where $\Delta_{\mathbf{k}}^{\alpha}$ possesses the same phase.

A further test of the phase sensitivity of $\Delta g(\mathbf{q}, E, B)$ is by contrasting its form with that obtained for a nodal $s$-wave superconductor in which the phase of the superconducting order parameter is uniform. This implies that in the calculation of $\Delta g(\mathbf{q}, E, B)$, we replace $\Delta_{\mathbf{k}}^{\alpha,\beta}$ by its magnitude $\left|\Delta_{\mathbf{k}}^{\alpha,\beta}\right|$ in Eq. S25 – Eq. S29. The results for the same set of parameters as before [i.e., those used in Figs. S2(u) – (y)] are shown in Figs. S2(z6) – (z10). These results show a structure of $\Delta g(\mathbf{q}, E, B)$ that is qualitatively different from the experimental data shown in Figs. S2(z1) – (z5), clearly demonstrating the presence of a phase-dependent superconducting gap. This together with the nodal structure necessary to explain the BQPI data, provides strong evidence for a $d_{x^2-y^2}$-wave symmetry of the superconducting order parameter.

We note in passing that if the ratio $M_{cc}/U_{cc}$ is such that $\bar{g}(\mathbf{q}, E, B)$ and $\bar{g}(\mathbf{q}, E, B = 0)$ possess the same sign, one obtains from Eq. S34

$$\begin{aligned}\Delta g(\mathbf{q}, E, B) &= \zeta\{U_{cc}[\bar{g}_G(\mathbf{q}, E) - \bar{g}_F(\mathbf{q}, E)] + M_{cc}[\bar{g}_G(\mathbf{q}, E) + \bar{g}_F(\mathbf{q}, E)] \\ &\quad - U_{cc}[\bar{g}_G(\mathbf{q}, E) - \bar{g}_F(\mathbf{q}, E)]\} \\ &= \zeta M_{cc}[\bar{g}_G(\mathbf{q}, E) + \bar{g}_F(\mathbf{q}, E)] = \zeta \bar{g}_{mag}(\mathbf{q}, E) \end{aligned} \quad (S37)$$

and the PQPI spectrum thus directly reflects the form of the magnetic BQPI spectrum, which is not a direct probe for the phase of the SC gap.

### 2.3.2 Experimental study of PQPI

To study the field dependence of the scattering interference pattern, we image the differential conductance $g(\mathbf{r}, E)$ at zero field and B=3T with atomic resolution and register. In order to be able to compare the zero field and in-field maps, the two measurements have been taken at the exact same location, using the same tip and scanning parameters. The first step in analyzing the data is to use a drift correction algorithm (19) to perfectly match and align the measurements as e.g. thermal drift will be slightly different however careful the measurements have been performed. Then, $g(\mathbf{q}, E, B)$, the square root of the power spectral density of each image is determined. To increase signal-to-noise, we fourfold symmetrize $g(\mathbf{q}, E, B)$, along the principle directions. The drift corrected $g(\mathbf{r}, E, B)$ and corresponding four-fold symmetrized $g(\mathbf{q}, E)$, are shown in Fig. S3 for five energies within the superconducting gap. To enable easy comparison, pairs of $g(\mathbf{r}, E, B)$, and $g(\mathbf{q}, E, B)$, at identical $E$ are shown using an identical color scale.

Then, we subtract the zero field from the in-field $g(\mathbf{q}, E, B)$ to obtain $\Delta g(\mathbf{q}, E, B)$. The experimental $\Delta g(\mathbf{q}, E = -0.5 \text{ meV}, B)$ and its histogram are plotted in Fig. S4(b) and Fig. S4(a), respectively. To clearly visualize the enhanced and suppressed scattering vectors, the histogram has been divided into three colors: enhanced vectors (blue), suppressed vectors (red), and the background (white) to which the bulk of the pixels is attributed, resulting in Fig. S4(c). The panels shown in Fig. 4 of the main text and in Fig. S2 and Fig. S4(d) are 3-pixels boxcar averages of the three valued images.

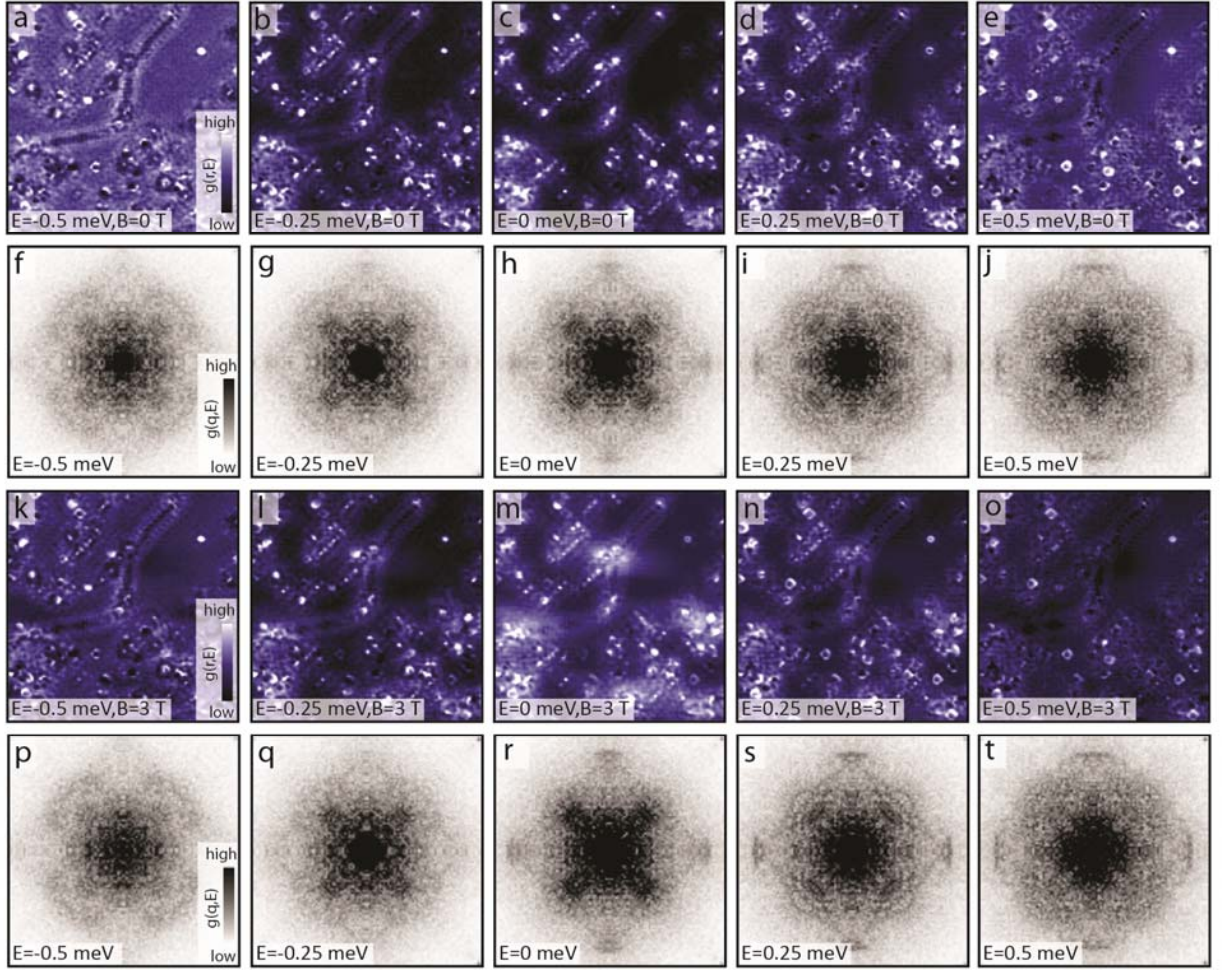

**Fig. S3.** Experimental PQPI taken at 250 mK (V=-10 meV, I=300 pA). On the exact same field of view (19x19 nm²), $g(\mathbf{r}, E, B)$ have been taken at 0 Tesla (a-e) and at 3 Tesla (k-o). The corresponding symmetrized $g(\mathbf{q}, E, B)$ are shown in (f-g) and (p-t) respectively (right- top corner is (2π,0)).

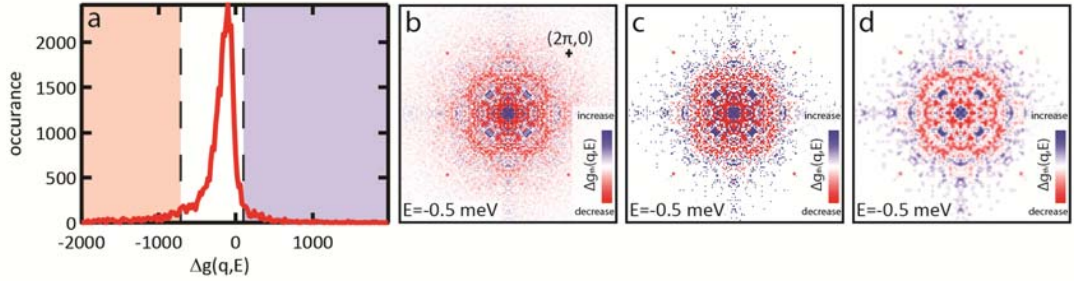

**Fig. S4**. Experimental PQPI analysis. The (a) histogram of the (b) difference of the symmetrized in field and zero-field $g(\mathbf{q}, E, B)$ is divided into three parts, where the bulk of the pixels (i.e. the background) is set to zero (white). The (c) three valued result is (d) boxcar averaged over three pixels.

## 3. Spin Excitations of CeCoIn$_5$: Spin-Resonance and the Spin-Lattice Relaxation Rate $1/T_1$

In superconductors with an unconventional symmetry, strong magnetic interactions can lead to the emergence of a magnetic resonance peak in the superconducting state. This peak was observed (20) in CeCoIn$_5$ in the imaginary part of the dynamical spin susceptibility at an energy of $E \approx 0.6$ meV. To describe the emergence of the magnetic resonance peak in CeCoIn$_5$, we compute the spin susceptibility in the random-phase approximation (RPA). This approach has previously been employed to successfully explain the existence of a resonance peak in the high-temperature superconductors (21-23) and heavy fermion materials (24,25).

Since the by far largest contribution to the magnetic susceptibility, $\chi$, arises from the magnetic $f$-moments, and since the energy and momentum position of the resonance are unaffected by contributions from the light $c$-bands, we neglect the latter and define $\chi$ in Matsubara $\tau$-space via

$$\chi(\mathbf{r} - \mathbf{r}', \tau) = \langle T_\tau \mathbf{S}_\mathbf{r}(\tau) \cdot \mathbf{S}_{\mathbf{r}'}(0) \rangle$$
$$= \frac{1}{2} \langle T_\tau S_\mathbf{r}^+(\tau) S_{\mathbf{r}'}^-(0) \rangle + \frac{1}{2} \langle T_\tau S_\mathbf{r}^-(\tau) S_{\mathbf{r}'}^+(0) \rangle + \langle T_\tau S_\mathbf{r}^z(\tau) S_{\mathbf{r}'}^z(0) \rangle$$
$$= \chi^\pm(\mathbf{r} - \mathbf{r}', \tau) + \chi^\mp(\mathbf{r} - \mathbf{r}', \tau) + \chi^{zz}(\mathbf{r} - \mathbf{r}', \tau) \quad \text{(S38)}$$

where $\mathbf{S_r}$ is the spin-1/2 operator describing the moments of the $f$-electrons.

The retarded, non-interacting spin susceptibility in the superconducting state for a single spin degree of freedom, $\chi_0^{SC}(\mathbf{q}, \omega)$, is then given by

$$\chi_0^{SC}(\mathbf{q},\omega) = -\frac{1}{2N}\sum_{\mathbf{k}}\sum_{i,j=\alpha,\beta}\zeta_{\mathbf{k},i}^2\zeta_{\mathbf{k+q},j}^2\left\{\left(1+\frac{E_{\mathbf{k}}^i E_{\mathbf{k+q}}^j + \Delta_{\mathbf{k}}^i\Delta_{\mathbf{k+q}}^j}{\Omega_{\mathbf{k}}^i\Omega_{\mathbf{k+q}}^j}\right)\frac{n_F(\Omega_{\mathbf{k}}^i) - n_F(\Omega_{\mathbf{k+q}}^j)}{\omega + i\delta + \Omega_{\mathbf{k}}^i - \Omega_{\mathbf{k+q}}^j}\right.$$
$$\left.+\left(1-\frac{E_{\mathbf{k}}^i E_{\mathbf{k+q}}^j + \Delta_{\mathbf{k}}^i\Delta_{\mathbf{k+q}}^j}{\Omega_{\mathbf{k}}^i\Omega_{\mathbf{k+q}}^j}\right)\frac{\left(\Omega_{\mathbf{k}}^i+\Omega_{\mathbf{k+q}}^j\right)}{(\omega+i\delta)^2 - \left(\Omega_{\mathbf{k}}^i+\Omega_{\mathbf{k+q}}^j\right)^2}\left[1-n_F(\Omega_{\mathbf{k}}^i)-n_F\left(\Omega_{\mathbf{k+q}}^j\right)\right]\right\}$$

(S39)

where

$$\zeta_{\mathbf{k},i}^2 = \begin{cases} w_{\mathbf{k}}^2 & \text{if } i = \alpha \\ x_{\mathbf{k}}^2 & \text{if } i = \beta \end{cases}$$

(S40)

and $\delta = 0^+$. Note that in the momentum sum in Eq. S39, $\Delta_{\mathbf{k}}^{\alpha\beta} \neq 0$ only for those momentum states within the Debye energy of the Fermi energy.

Starting from a bare (static) magnetic interaction between the $f$-moments in the spin-flip channel (similar to the full interaction given in Eq. S14)

$$H_{sf} = \frac{1}{2N}\sum_{\mathbf{k,p,q}} I_0(\mathbf{q}) f_{\mathbf{k+q},\uparrow}^\dagger f_{\mathbf{p-q},\downarrow}^\dagger f_{\mathbf{k},\downarrow} f_{\mathbf{p},\uparrow} = \frac{1}{N}\sum_{\mathbf{k,p,q}} \bar{I}_0(\mathbf{q}) f_{\mathbf{k+q},\uparrow}^\dagger f_{\mathbf{p-q},\downarrow}^\dagger f_{\mathbf{k},\downarrow} f_{\mathbf{p},\uparrow}$$

(S41)

with $\bar{I}_0(\mathbf{q}) = I_0(\mathbf{q})/2$ we obtain within the random phase approximation for the full transverse susceptibility

$$\chi_{SC}^{\pm}(\mathbf{q},\omega) = \frac{1}{2}\frac{\chi_0^{SC}(\mathbf{q},\omega)}{1+\bar{I}_0(\mathbf{q})\chi_0^{SC}(\mathbf{q},\omega)}$$

(S42)

To obtain $I_0(\mathbf{q})$, we first note that the magnetic interaction $I(\mathbf{q})$ in Eq. S13 is the fully renormalized interaction since it was extracted from the experimental QPI data. It can be related to $I_0(\mathbf{q})$ within the random phase approximation in the normal state via

$$\left[\frac{I(\mathbf{q})}{2}\right]^{-1} = [\bar{I}(\mathbf{q})]^{-1} = [\bar{I}_0(\mathbf{q})]^{-1} + \text{Re}\chi_0^N(\mathbf{q},\omega) \tag{S43}$$

Here, we relate $\bar{I}_0(\mathbf{q})$ and $\bar{I}(\mathbf{q})$ in the normal state since $I(\mathbf{q})$ was computed from Eq. S2b using the normal state Greens functions. Since $\text{Re}\chi_0^N(\mathbf{q},\omega)$ in the normal state is only weakly frequency dependent at small frequencies, as shown in Fig. S5a for $\mathbf{q} = (\pi,\pi)$, we set $\text{Re}\chi_0^N(\mathbf{q},\omega) \approx \text{Re}\chi_0^N(\mathbf{q},\omega=0) = \chi_0^N(\mathbf{q},\omega=0)$ and thus obtain for the bare interaction

$$[\bar{I}_0(\mathbf{q})]^{-1} = [\bar{I}(\mathbf{q})]^{-1} - \chi_0^N(\mathbf{q},\omega=0) \tag{S44}$$

We note in passing that replacing $\chi_0^N(\mathbf{q},\omega=0)$ by $\chi_0^{SC}(\mathbf{q},\omega=0)$ in Eq. S44, and thus considering the renormalization of the interaction in the superconducting state, has only a weak effect on the form of $I_0(\mathbf{q})$.

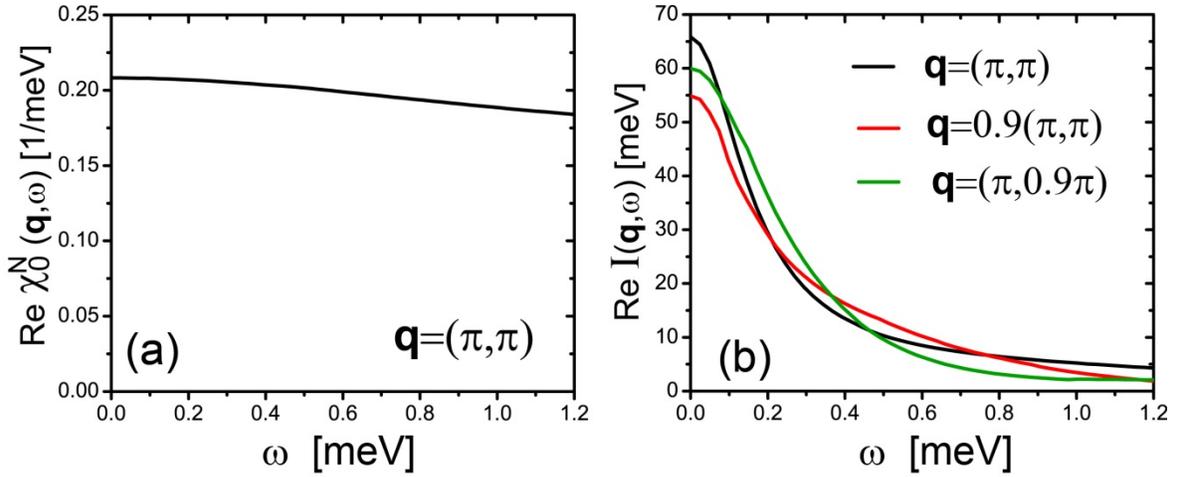

**Fig. S5.** (a) Re $\chi_0^N(\mathbf{q},\omega)$ in the normal state as a function of frequency, $\omega$, at $T = 0$. It exhibits only a weak dependence on $\omega$. (b) Re $I(\mathbf{q},\omega)$ in the normal state as a function of frequency, $\omega$, at $T = 0$, as obtained from Eq. S50. Shown are the results for three momenta near $\mathbf{q} = (\pi,\pi)$.

Inserting Eq. S44 into Eq. S42 then yields

$$\chi_{SC}^\pm(\mathbf{q},\omega) = \frac{1}{2} \frac{\chi_0^{SC}(\mathbf{q},\omega)[[\bar{I}(\mathbf{q})]^{-1} - \chi_0^N(\mathbf{q},\omega=0)]}{[\bar{I}(\mathbf{q})]^{-1} + [\chi_0^{SC}(\mathbf{q},\omega) - \chi_0^N(\mathbf{q},\omega=0)]} \tag{S45}$$

The imaginary part of the dynamical susceptibility, Eq. S45, exhibits a magnetic resonance peak at $\mathbf{q} = (\pi, \pi)$ and $E \approx 0.6$ meV, as shown in Fig. 4F of the main text, in very good agreement with the experimental findings (20) reproduced in Fig. 4G.

Moreover, we note that the position of the resonance peak, whose position in energy determined via

$$[\bar{I}(\mathbf{q})]^{-1} - \text{Re}[\chi_0^{SC}(\mathbf{q}, \omega) - \chi_0^N(\mathbf{q}, 0)] = 0 \tag{S46}$$

is unaffected by contributions to the spin susceptibility from the $c$-electron band, since the magnetic interaction occurs only within the $f$-electron band.

Finally, the spin lattice relaxation rate is given by

$$\frac{1}{T_1} = \frac{k_B T}{2\hbar} (\hbar^2 \gamma_n \gamma_e)^2 \frac{1}{N} \sum_{\mathbf{q}} A(\mathbf{q}) \lim_{\omega \to 0} \frac{2\,\text{Im}\chi_{SC}^{\pm}(\mathbf{q}, \omega)}{\omega} \tag{S47}$$

where $A(\mathbf{q})$ is the hyperfine coupling constant, and the factor 2 on the r.h.s of Eq. S45 reflects the two spin degrees of freedom contributing to the total transverse susceptibility. Since the microscopic form of $A(\mathbf{q})$ is unknown, we will consider a direct hyperfine constant only, in which case $A(\mathbf{q})$ is momentum independent, i.e., $A(\mathbf{q}) = A_0$.

To compute the temperature dependence of $1/T_1$, we calculate the temperature dependent superconducting gap, $\Delta_{\mathbf{k}}^{\alpha,\beta}(T)$, from superconducting gap equation, Eq. S20, and then use its form in computing $\chi_0^{SC}(\mathbf{q}, \omega, T)$ in Eq. S39. To obtain $\chi_{SC}^{\pm}(\mathbf{q}, \omega, T)$ in Eq. S41, which enters the calculation of $1/T_1$ in Eq. S45, it is now necessary to compute the bare interaction, $I_0(\mathbf{q})$, for the entire magnetic Brillouin zone for $T = 0$. Moreover, since we compute $1/T_1$ over a larger temperature range, and to avoid any spurious effects in the calculation of $\chi_{SC}^{\pm}(\mathbf{q}, \omega)$ at small momenta $\mathbf{q}$, we now relate $I_0(\mathbf{q})$ and $I(\mathbf{q})$ in the superconducting state via

$$[\bar{I}_0(\mathbf{q})]^{-1} = [\bar{I}(\mathbf{q})]^{-1} - \chi_0^{SC}(\mathbf{q}, \omega = 0, T = 0) \tag{S48}$$

such that the full spin-susceptibility in the entire Brillouin zone for $T < T_c$ is given by

$$\chi_{SC}^{\pm}(\mathbf{q}, \omega, T) = \frac{1}{2} \frac{\chi_0^{SC}(\mathbf{q}, \omega, T)[[\bar{I}(\mathbf{q})]^{-1} + \chi_0^{SC}(\mathbf{q}, \omega = 0, T = 0)]}{[\bar{I}(\mathbf{q})]^{-1} + [\chi_0^{SC}(\mathbf{q}, \omega, T) - \chi_0^{SC}(\mathbf{q}, \omega = 0, T = 0)]} \tag{S49}$$

The resulting temperature dependence of $1/T_1$, is shown in Fig. 4D of the main text.

The temperature dependence of $1/T_1$ is in general determined by a superposition of contributions from inter- and intra-band scattering. As a result, one recovers the expected $d$-wave result $1/T_1 \sim T^3$ only for $k_B T$ much smaller than the smallest superconducting gap (on the $\beta$-FS). In the experimentally relevant intermediate temperature regime where $k_B T$ is larger than (or comparable to) the small superconducting gap on the $\beta$-FS, but smaller than the large gap on the $\alpha_1$-FS, the temperature dependence of $1/T_1$ is given by a superposition of different power-laws. Nevertheless, we find that in this regime, the theoretical results for $1/T_1$ can be well fitted by an approximate power-law $1/T_1 \sim T^\alpha$ with $\alpha \approx 2.5$, as shown in Fig. 4D of the main text.

## 4. Debye Energy, Spin Fluctuation Spectrum and Renormalized Interactions

While $\omega_D$ was introduced as a phenomenological parameter in section S2, the question naturally arises of whether it can be related to the energy scale of the spin fluctuation spectrum in the normal state; the latter can be considered the equivalent of the weak-coupling $\omega_D$ in a strong coupling theory of superconductivity. To investigate this question, we consider the energy dependence of $\mathrm{Im}\chi(\mathbf{q}, E)$ for $\mathbf{q} = (\pi, \pi)$ as shown in Fig. 4F of the main text. It exhibits a maximum around $E = 0.2$ meV, and approximately 50% of its total spectral weight is located below $E = 1$ meV, consistent with an $\omega_D$ in the range of 0.5 to 1 meV.

In addition, we can consider the frequency dependence of the full RPA interaction

$$I(\mathbf{q}, \omega) = \frac{I_0(\mathbf{q})}{1 + \frac{I_0(\mathbf{q})}{2} \chi_0^N(\mathbf{q}, \omega)} \qquad (S50)$$

as obtained using the bare interaction, $I_0(\mathbf{q})$ in Eq. S44.

In Fig. S5b, we present the frequency dependence of the real part of $I(\mathbf{q}, \omega)$ (as obtained from Eq. S50 using $I_0(\mathbf{q})$ obtained in Sec. S3) for several momenta near $\mathbf{q} = (\pi, \pi)$. The interaction decreases on the scale of $\omega \approx 0.3 - 0.5$ meV, a scale that is also (at least qualitatively) consistent with that of the Debye energy, $\omega_D$. The qualitative consistency of $\omega_D$ we had predicted in section S2.1, with the energy scale of the spin fluctuations, as well as that of the renormalized RPA interactions, might provide the first clue to its microscopic origin. A more detailed investigation of $I(\mathbf{q}, \omega)$, its relation to the Debye energy, and the resulting frequency dependent hopping parameter, $t_{fi}$, requires the development of a theory that accounts for the dynamical aspects of both the slave boson and the magnetic correlations which is beyond the scope of the present study.

**References**

1  Coleman P (1984) New approach to the mixed-valence problem. *Phys. Rev. B* 29(6): 3035.
2  Millis AJ, Lee PA (1987) Large-orbital-degeneracy expansion for the lattice Anderson model. *Phys. Rev. B* 35(7): 3394.
3  Newns DM, Read, N (1987) Mean-field theory of intermediate valence/heavy fermion systems. *Adv. Phys.* 36(6): 799.
4  Kotliar G, Liu J (1988) Superexchange mechanism and *d*-wave superconductivity. *Phys. Rev. B* 38(7): 5142.
5  Read N, Newns DM (1983) On the solution of the Coqblin-Schreiffer Hamiltonian by the large-N expansion technique. *J. Phys. C* 16(17): 3273.
6  Coleman P (1983) 1/N expansion for the Kondo lattice. *Phys. Rev. B* 28(9): 5255.
7  Bickers NE (1987) Review of techniques in the large-N expansion for dilute magnetic alloys. *Rev. Mod. Phys.* 59(4): 845.
8  Senthil T, Vojta M, Sachdev S (2004) Weak magnetism and non-Fermi liquids near heavy-fermion critical points. *Phys. Rev. B* 69(3): 035111.
9  Booth CH, et al. (2011) Electronic structure and *f*-orbital occupancy in Yb-substituted CeCoIn$_5$. *Phys. Rev. B* 83(23): 235117.

10  Yuan T, Figgins J, Morr DK (2012) Hidden order transition in $URu_2Si_2$: Evidence for the emergence of a coherent Anderson lattice from scanning tunneling microscopy. *Phys. Rev. B* 86(3): 035129.

11  Abanov Ar, Chubukov AV, Finkel'stein AM (2001) Coherent vs. incoherent pairing in 2D systems near magnetic instability. Europhys. Lett., 54 (4): 488.

12  Allan MP, et al. (2013) Imaging Cooper pairing of heavy fermions in CeCoIn5. *Nature Physics* 9(8): 468-473.

13  Movshovich R, et al. (2001) Unconventional superconductivity in $CeIrIn_5$ and $CeCoIn_5$: Specific heat and thermal conductivity studies. *Phys. Rev. Lett.* 86(22): 5152.

14  Maiti S, Chubukov AV (2011) Relation between nodes and $2\Delta/T_c$ on the hole Fermi surface in iron-based superconductors. *Phys. Rev. B* 83(22): 220508.

15  Dolgov OV, Mazin II, Parker D, Golubov AA (2009) Interband superconductivity: Contrasts between Bardeen-Cooper-Schrieffer and Eliashberg theories. *Phys. Rev. B* 79(6): 060502.

16  Musaelian KA, Betouras J, Chubukov AV, Joynt R (1996) Mixed-symmetry superconductivity in two-dimensional Fermi liquids. *Phys. Rev. B* 53(6): 3598.

17  Hanaguri T, et al. (2009) Coherence factors in a high-$T_c$ cuprate probed by quasi-particle scattering off vortices. *Science* 323(5916): 923-926.

18  Hanaguri T, Niitaka S, Kuroki K, Takagi H (2010) Unconventional *s*-wave superconductivity in Fe(Se,Te). *Science* 328(5977): 474.

19  Lawler MJ, et al. (2010) Intra-unit-cell electronic nematicity of the high-$T_c$ copper-oxide pseudogap states. *Nature* 466(7304): 347–351.

20  Stock C et al. (2008) Spin resonance in the *d*-wave superconductor CeCoIn5. *Phys. Rev. Lett.* 100(8): 087001.

21  Mazin II, Yakovenko VM (1995) Neutron scattering and superconducting order parameter in $YBa_2Cu_3O_7$. *Phys. Rev. Lett.* 75(22): 4134.

22  Liu DZ, Zha Y, Levin K (1995) Theory of neutron scattering in the normal and superconducting states of $YBa_2Cu_3O_{6+x}$. *Phys. Rev. Lett.* 75(22): 4130.

23  Millis A, Monien H (1996) Bilayer coupling in the yttrium-barium family of high-temperature superconductors. *Phys. Rev. B* 54(22): 16172.

24  Eremin I, Zwicknagl G, Thalmeier P, Fulde P (2008) Feedback spin resonance in superconducting $CeCu_2Si_2$ and $CeCoIn_5$. *Phys. Rev. Lett.* 101(18): 187001.
25  Chubukov AV, Gorkov LP (2008) Spin resonance in three-dimensional superconductors: the case of $CeCoIn_5$. *Phys. Rev. Lett.* 101(14): 147004.